\documentclass[journal]{IEEEtran}
\usepackage{graphicx}
\usepackage{caption}
\usepackage{hyperref}
\usepackage{tabularx}
\usepackage{enumitem}

\begin{document}

\title{Learning Outcomes, Assessment, and Evaluation\\in Educational Recommender Systems:\\A Systematic Review}

\author{Nursultan Askarbekuly$^{1,2}$, Ivan Luković$^2$\\
\thanks{\\
$^1$ Faculty of Computer Science and Engineering, Innopolis University, Innopolis, Russia, e-mail: n.askarbekuly@innopolis.university\\
$^2$ Faculty of Organizational Sciences, The University of Belgrade, Belgrade, Serbia, e-mails: na20225034@student.fon.bg.ac.rs, ivan.lukovic@fon.bg.ac.rs}
}

\markboth{Journal X}%
{Askarbekuly and Luković: Systematic Review of Recommender Systems in E-learning}

\maketitle

\begin{abstract}
In this paper, we analyse how learning is measured and optimized in Educational Recommender Systems (ERS). In particular,  we examine the target metrics and evaluation methods used in the existing ERS research, with a particular focus on the pedagogical effect of recommendations. While conducting this systematic literature review (SLR), we identified 1395 potentially relevant papers, then filtered them through the inclusion and exclusion criteria, and finally selected and analyzed 28 relevant papers. Rating-based relevance is the most popular target metric, while less than a half of papers optimize learning-based metrics. Only a third of the papers used outcome-based assessment to measure the pedagogical effect of recommendations, mostly within a formal university course. This indicates a gap in ERS research with respect to assessing the pedagogical effect of recommendations at scale and in informal education settings.
\end{abstract}


\IEEEpeerreviewmaketitle

\section{Introduction}
\label{sec:introduction}
\IEEEPARstart{R}{ecommender} systems (RS) have been a subject of extensive research and development for the past two decades. Initially RS gained prominence in entertainment and e-commerce, and then have also been widely used in the education \cite{rivera2018recommendation, drachsler2015panorama, urdaneta2021recommendation}.

Entertainment, e-commerce, and education have their distinct use cases and goals, and it is reflected in how the recommendations are made and what metrics are optimized within each domain. In entertainment, RS were initially aimed at improving the user ratings \cite{bennett2007netflix}, and then the focus shifted to user engagement and retention \cite{knijnenburg2012explaining, maslowska2022role}. In e-commerce, recommendations are made to increase sales and probability of having return customers \cite{xiao2007commerce}. While in education, recommender systems mainly aim to increase learning through personalizing educational resources \cite{garcia2013educational}.

% A learner's interaction with a recommended item is not conclusive
Like any other software, an educational recommender system (ERS) requires an appropriate evaluation method aligned with its business goals \cite{gqm}. Yet learning is harder to measure than revenue or engagement, as ERS address a more complex use case, where the recommended item itself does not represent the end-goal. Instead, a learner's engagement with a recommended activity should promote learning \cite{urdaneta2021recommendation}.

There is no consensus on what is the best way to assess learning from the pedagogical point of view \cite{learning_measurement}, however one prominent approach to define learning is through the achievement of the intended learning outcomes \cite{europeanQualification, biggs2012student}. Adopting this outcome-based approach, we believe the evaluation of educational recommender systems should be rooted in outcome-based assessment \cite{askarbekuly2020combining}.

This systematic literature review (SLR) explores the existing research on educational recommended systems (ERS). The \emph{objective} of this SLR is to investigate \emph{how existing ERS increase learning}. Thus, we review what target metrics are optimized in existing research on ERS, what evaluation methods they use, and how the pedagogical effect of recommendations is measured from the outcome-based point of view.

The \emph{purpose} behind the SLR is to build a scalable outcome-based educational recommender system for the informal e-learning domain. This system's goal is to automate  the process of personalizing and evaluating educational trajectories within an informal learning domain. This includes the choice of learning outcomes for a particular learner, recommendation of appropriate learning activities, and evaluation of the effectiveness of recommendations through outcome-based assessment. Such automation can be of value to educators and developers of educational software, as it tackles the non-trivial task of automatically determining the best educational trajectory in informal domain at scale.

Section \ref{sec:background} provides the necessary background and mentions related work. Section \ref{sec:methodology} describes our review methodology. In Section \ref{sec:results}, we report the results, and then discuss them in Section \ref{sec:discussion}. Lastly, we conclude the article in Section \ref{sec:conclusion}. For a fast reading of this paper, please see subsections \ref{subsec:rqs} and \ref{subsec:analysis_criteria} for research questions and analysis criteria, then subsection \ref{subsec:key_results} for the key results.

\section{Background and Related Work}
\label{sec:background}

In this section, we set out to give a reader an overview of all necessary background knowledge on the topic of our review. We describe the general theory behind recommender systems, then analyze the previous systematic reviews on education recommenders, and identify the gap that this work attempts to fill. We conclude with the subsection on otucome based assessment.

\subsection{Recommender Systems}

Recommender systems (RS) are a significant subset of information filtering systems. They attempt to recommend relevant items by predicting user preferences or ratings. Traditionally, RS can be categorized into collaborative filtering (CF), content-based filtering (CBF), knowledge-based filtering (KB) and their hybridizations \cite{burke2002hybrid}. In addition to that, some modern recommender systems use approaches based on Machine Learning (ML).

Collaborative filtering (CF) methods are based on collecting and analyzing users’ interaction behaviors with the items, and predicting what users will like based on how the similarities in ratings between users. Generally, there are two types of collaborative filtering: user-based (UBCF) and item-based (IBCF). UBCF algorithms recommend items by finding users with ratings similar to the target user, and then recommend items that those users have rated highly. For example, in an educational app, if two students have an intersection of activities on which they performed well, and one of them performs well on a new activity, the system might recommend it to the other student. On the contrary, IBCF models focus on calculating the item-item similarities through calculating the intersection of user ratings for the items. For example, in an educational app, if activities A,B, and C have been highly rated by the same set of users, and a new user likes item A, then the system might recommend items B and C. Since all recommendations are based on the user ratings, the advantage of CF is that they do not require any item metadata or knowledge about the content. However, such systems suffer from the cold-start problem, in cases where there are no available ratings for new items or new users.

Content-based filtering (CBF) methods use specific information about each item to provide recommendations. These could be information describing the content of the items, like the genre, the director, or the actor of a movie. Similarly to IBCF, the system recommends items similar to those the user preferred in the past, however the key difference is that the similarity is calculated based on the metadata instead of the user ratings. In an educational app, if a student shows interest in an activity marked with 'Mathematics' tag, the system will recommend more content related to 'Mathematics'. The main difficulty of such systems is that they require the metadata, which is often difficult or expensive to generate. CBF systems can also suffer from over-specialization, where they only recommend items similar to those the user has already rated.

Knowledge-based filtering (KB) systems usually contain some knowledge of the user needs and how items relate to such needs. They allow to specify constraints and nuanced relations between items and users, which can be useful in certain fields. In a way, KB is similar to expert systems \cite{fernandez2009challenges} in that it recommends items based on specific rules and knowledge of the domain. An example of KB recommender is that if a user on an educational app wants to "prepare for a data science job", the system might know that certain items and subjects are a prerequisite to the job, and then recommend the items also taking the user background knowledge into account. A disadvantage of KB is that it often requires explicit user inputs and specification of relations between the modelled entities.

The main difference between the three methods lies in what they consider for recommendations. CF considers the behavior of users and the interactions between users and items. It does not need to know anything about the items themselves. CBF considers descriptions of the items, and relies on past user behavior and item similarity to make recommendations. Lastly, KBF makes recommendations based on explicit knowledge about the items, the users, and the ways these two relate to each other.

Hybrid Systems: These systems combine the aforementioned methods in an attempt to avoid the limitations. For instance, a CF model might suffer from the "cold start" problem, where it's hard to recommend items to a new user because there's not enough data about them yet. CBF or KBF systems can help here by recommending items based on a few initial preferences set by the new user. CBF systems can suffer from over-specialization, and CF or KBF methods can help expand the user's interests based on ratings or knowledge of relations between the items and user's needs.

For a more comprehensive description of various techniques and their hybridizations, we recommend a foundational survey by Burke \cite{burke2002hybrid}, who analyzed existing recommendation techniques in terms of the input data and the algorithms, and highlighted general rules on when to use each technique, and examined the range of hybridization methods that have been proposed in the literature.

Apart from the traditional ones, other more sophisticated approaches to recommendation systems are also emerging. Context-Aware Recommender Systems \cite{review_context6189308} consider user's additional contextual information like time, location, or the device used when providing recommendations. This ability to adapt recommendations based on various contextual factors can potentially enhance the learning experience and lead to improved learning outcomes \cite{RAZA201984}. Deep Learning models \cite{deeplearning2022} can recognize complex patterns by training on large amounts of data. They are used to create recommender systems that can generate more thorough metadata by analyzing the content of items or take into account other factors, which are otherwise ignored by the more traditional systems. Reinforcement Learning models \cite{reinforcement2022} learn by trial and error. They can be used to create recommender systems that adapt to changes in user preferences and behavior over time.

\subsection{Reviews on Recommender Systems in Education}

There are several existing reviews of the recommender systems in education, each of them focusing on an aspect of the process. Drachsler et al \cite{drachsler2015panorama} analyzed recommender systems in Technology-Enhanced Learning (TEL) in the period from year 2000 to 2014, and  proposed a classification framework for RS in TEL. They noted that most RS can be described in terms of the learner model, domain model, and recommendation algorithm. They also made an important observation that the background knowledge and learner's goals should be part of the learner model. \cite{dwivedi2017recommender} made a concise overview of big data methods and tools used in recommender systems. Rivera et al \cite{rivera2018recommendation} conducted a general systematic mapping and categorized articles according to the educational context, RS approaches, use case, and evaluation of results. Urdaneta-Ponte et al \cite{urdaneta2021recommendation} also highlight that RS should be tailored to the domain, education having three of them: formal education, non-formal education, and informal education. A review by Verbert et al \cite{review_context6189308} outline the type of contextual information about the user that can serve as input in the recommendation process. Thongchotchat et al \cite{review_styles10022322} review how learning styles are used as part of the learner context in ERS. Da Silva et al \cite{da2023systematic} review educational recommender systems in regards to how recommendations have been produced, what are the most common methods. They touch on evaluation and note the prevalence of accuracy-based metrics.

However, none of the aforementioned reviews focus on the role of learning outcomes in the recommendation process, nor discuss the outcome-based assessment and evaluation in ERS. Furthermore, we found no systematic review comprehensively examining what is optimized in ERS, i.e. the target metrics and evaluation methods together with the educational context and supported tasks.

\subsection{Outcome-based Evaluation in ERS}

Any recommender system deployed in a domain should have evaluation metrics reflecting the unique goals of this system and broader goals of the domain \cite{burke2002hybrid, gqm}. Standard recommender systems, such as those used in e-commerce or entertainment, often prioritize the following metrics:
\begin{itemize}
    \item Click-Through Rate (CTR): Measures the ratio of users who click on a recommended item to the number of total users who view the recommendation, indicating immediate interest but not necessarily long-term engagement or satisfaction.
    \item Conversion Rate: In e-commerce, this metric measures the percentage of recommendations that result in a purchase, focusing on direct revenue impact.
    \item Engagement and Retention: indicates the system's success in maintaining users' interest and facilitating ongoing usage.
    \item Accuracy Metrics: Such as precision, recall, and F1 score, are common in many types of recommender systems.
    \item User Satisfaction and Diversity: reflect content enjoyment or variety from a broad user-centric perspective.
\end{itemize}

These standard metrics can be relevant to educational contexts, though their interpretation may differ. In addition to that, ERS should use more outcome-based metrics focusing on pedagogical effectiveness \cite{drachsler2015panorama, garcia2013educational}, such as:
\begin{itemize}
    \item Learning Gain or Improvement: the increase in knowledge or skills of learners after engaging with the recommended educational resources, e.g. assessed through pre-tests and post-tests.
    \item Concept Mastery and Skill Acquisition: Uses assessments to evaluate how well learners have mastered specific concepts or acquired particular skills recommended by the system.
\end{itemize}

This learning-centered metrics are usually derived from the learning outcomes (LOs). This outcome-based approach became de-facto European union standard for learning and teaching \cite{europeanQualification}. The main idea behind outcome-based learning is that it is important to ask what competences, knowledge, and skills the learner gains as the result of the educational process. Biggs \cite{biggs2012student} also adheres to this approach, and proposes the idea of constructive alignment (CA). CA states that both learning activities and assessment should be aligned with the intended learning outcomes. At, its essence, an intended learning outcome is a goal that the educator plans for a learner. In our earlier works \cite{askarbekuly2021building, askarbekuly2020combining}, we have demonstrated that it is possible to align intended outcomes, learning activities, and learning assessment when building an educational product.

To summarize, we assume that educational recommender systems should facilitate learning, i.e. promote the achievement of learning outcomes.  To check this assumption, we review existing primary ERS research to check what educational recommenders actually try to impact. By extension, we examine how existing ERS assess and verify that the intended outcomes are achieved.

\section{Methodology}
\label{sec:methodology}

In this section, we describe the process of our review. Our work uses the SLR methodology proposed by Kitchenham \cite{kitchenham2004procedures} and partially borrows the paper structure from Farina et al \cite{farina2022technologies}. We start from stating the research questions and discuss their relevance to the overall research objective. Then we describe the analysis criteria based on the research questions. Lastly, we protocol our search strategy including the databases, search queries, and inclusion/exclusion criteria used in the review.

\subsection{Research questions}
\label{subsec:rqs}

Our research questions follow from the objective and purpose of this SLR (Section \ref{sec:introduction}). \emph{The objective} of this SLR is to investigate how educational recommender systems (ERS) model and optimize learning. \emph{The purpose} behind the objective is to build an outcome-based ERS for the informal e-learning domain, including the identification of learning outcomes for a particular learner, recommendation of learning activities, and evaluation of the effectiveness of recommendations through outcome-based assessment.

We translated the objective into three research questions keeping in mind the overall purpose of our investigation:

\begin{itemize}[label={}, left=5pt]
    \item \emph{RQ1: What target metrics are optimized in educational recommender systems (ERS)?}
    \item \emph{RQ2: What evaluation methods are used for measuring the effectiveness of ERS?}
    \item \emph{RQ3: How do educational recommender systems implement outcome-based assessment?}
\end{itemize}

\emph{RQ1} checks the general assumption, embedded in the objective, that ERS actually aim to optimize learning. It also serves as the basis to further examine how ERS optimize learning outcomes. \emph{RQ2} explores how researchers evaluate the effectiveness of ERS given their target metrics. Lastly, \emph{RQ3} investigates how outcome-based assessment is implemented in ERS, and how they relate to the target metric and evaluation methods.

Answering these questions will instruct us on how to measure the effectiveness of recommendations against learning outcomes as the target metric. This is coherent with our overall outcome-based approach and the purpose of this SLR described above.

\subsection{Search strategy}
\label{subsec:search_strategy}

Based on the three research questions we developed a search strategy for selecting the relevant papers. For our search, we have chosen the following databases popular in the Computer Science field: IEEE Xplore, ACM Digital Library, and SpringerLink. We selected IEEE, ACM, and SpringerLink as they are among the most significant and reputable sources for research in the field of Computer Science. The research published in these databases is typically of acceptable quality and peer-reviewed. We did not include other databases such as DBLP, Scopus, or Web of Science, as there is likely to be an overlap with the papers indexed in IEEE, ACM, and SpringerLink. We also excluded databases like Google Scholar and arXiv, as they include a wide range of not peer-reviewed content, including theses, white papers, and reports, that might not be as relevant or as rigorously vetted.

Table \ref{tab:databases} reviews the databases used for this SLR in terms of their advantages and disadvantages. The most important feature of all the three databases is that they are focused on technology and have sound peer-review process to guarantee the quality of publications.

\begin{table}[h!]
\centering
\caption{Advantages and Disadvantages of Each Database}
\begin{tabular}{|p{1.5cm}|p{4cm}|p{2cm}|}
\hline
\textbf{Database} & \textbf{Advantages} & \textbf{Disadvantages} \\
\hline
IEEE Xplore &
\begin{itemize}
\item High-quality, peer-reviewed content.
\item Specializes in technical literature in electrical engineering, computer science, and electronics.
\item Advanced search features.
\end{itemize} &
Most of the content is paid.\\
\hline
SpringerLink &
\begin{itemize}
\item Offers access to millions of scientific documents from journals, books, series, protocols, and reference works.
\item Content from various disciplines.
\item High-quality, peer-reviewed content.
\end{itemize} &
Full access usually requires a subscription.\\
\hline
ACM Digital Library &
\begin{itemize}
\item Comprehensive collection of ACM's publications.
\item Provides access to a broad range of computer science material.
\item High-quality, peer-reviewed content.
\end{itemize} &
Full access usually requires a subscription.\\
\hline
\end{tabular}
\label{tab:databases}
\end{table}

Given the topic of our SLR, we have determined the "educational recommender systems" to be an appropriate keyphrase.

\begin{table}[h!]
\centering
\caption{Keywords and synonyms}
\label{tab:keywords}
\begin{tabular}{|l|l|}
\hline
\textbf{keyword} & \textbf{synonyms} \\
\hline
educational & learning, training, teaching, tutoring \\
\hline
recommender & intelligent, adaptive \\
\hline
system & - \\
\hline
\end{tabular}
\end{table}

After a set of trials and preliminary exploration, we also added synonyms to "educational" and "recommender". As the synonyms of "educational", we used "learning", "training", "teaching", and "tutoring". For "recommender", we used "intelligent" and "adaptive" (Table \ref{tab:keywords}).

Thus, we were able to also cover intelligent tutoring systems and adaptive learning system, which both recommend and personalize resources and aid for learners. Using the search queries with boolean operators, we have arrived at the following search results:

\begin{itemize}
    \item IEEE Xplore:\\
    ("Document Title":educat* OR "Document Title":learn* OR "Document Title":teach* OR "Document Title":train* OR "Document Title":tutor*) AND (("Document Title":recommend* OR "Document Title":intelligent* OR "Document Title":adapt*) NEAR/1 ("Document Title":system))
    \item ACM Digital Library:\\
    Title: (educat* OR learn* OR teach* OR train* OR tutor*) AND (recommend* OR intelligent* OR adapt*) AND (system)
    \item SpringerLink:\\
    For this database, we had to conduct four separate searches as it only accepts basic unnested querries\\
    Title: (educat* AND recommend* system)\\
    Title: (learn* AND recommend* system)\\
    Title: (teach* AND recommend* system)\\
    Title: (train* AND recommend* system)\\
    Title: (tutor* AND recommend* system)\\
    Title: (educat* AND intelligent* system)\\
    Title: (learn* AND intelligent* system)\\
    Title: (teach* AND intelligent* system)\\
    Title: (train* AND intelligent* system)\\
    Title: (tutor* AND intelligent* system)\\
    Title: (educat* AND adapt* system)\\
    Title: (learn* AND adapt* system)\\
    Title: (teach* AND adapt* system)\\
    Title: (train* AND adapt* system)\\
    Title: (tutor* AND adapt* system)\\
\end{itemize}

To filter the relevant papers we have determined Inclusion and Exlusion criteria. To be included, a paper must satisfy all of the following Inclusion Criteria (IC).
\begin{enumerate}
    \item IC1: the work is written in readable English without major understandability issues AND
    \item IC2: the work describes a recommender system in the context of education
\end{enumerate}

A work would be excluded if it met any of the following Exclusion Criteria (EC):
\begin{enumerate}
    \item EC1: the work is similar to a later work by the same authors OR
    \item EC2: the work is a conference proceeding
    \item EC3: the work is grey literature (e.g. technical report, dissertation, a newsletter, a standard).
\end{enumerate}

A potential limitation of our study is the use of only three databases and the focus on the journal publications, driven by our intention to select the most relevant and high-quality papers. In the further work, it could be possible to extend our sources to other databases and conference publications, if we have research hypotheses and goals justified by our initial approach.

\subsection{Analysis criteria}
\label{subsec:analysis_criteria}

To answer the research questions, we defined analysis criteria (Table \ref{tab:specific_analysis_criteria}) to investigate in the selected papers.

\begin{table}[htbp]
    \centering
    \caption{Analysis criteria and corresponding research questions}
    \begin{tabular}{clc}
        \textbf{ID} & \textbf{Criterion} & \textbf{Research Question} \\
        AC1 & Target Metrics & RQ1 \\
        AC2 & Evaluation Methods & RQ2 \\
        AC3 & Inclusion of Learning Outcomes & RQ3 \\
        AC4 & Assessment of Learning Outcomes & RQ3 \\
        \label{tab:specific_analysis_criteria}
    \end{tabular}
\end{table}

\emph{AC1} examines what each of the ERS optimizes and aims to increase. \emph{AC2} reviews how target metrics are evaluated. \emph{AC1} and \emph{AC2} directly correspond and respectively address \emph{RQ1} and \emph{RQ2}. Then through \emph{AC3} and \emph{AC4}, we examine whether and how the reviewed systems assess the achievement of outcomes (\emph{RQ3}). \emph{AC3} also reviews the role of learning outcomes in the recommendation process and how they relate to target metrics. While \emph{AC4} can help us find techniques for conducting the assessment efficiently. Overall, these criteria enable us to identify outcome-oriented recommenders and to explore how they model learning and assessment.

In addition to the criteria specific to our research questions, we also explored more general ERS attributes such as the context in which the system is deployed,    supported tasks, end-users, and recommendation method used. This aspects become especially important to examine in outcome-oriented recommenders.

To summarize the section, we derived research questions from the objective and purpose of the SLR, then defined search strategy to select relevant papers, and established the analysis criteria to answer the research questions. In the following section, we present search results, provide the general overview of the examined ERS, and then proceed to the main analysis criteria and corresponding research questions.

\section{Results}
\label{sec:results}

In this section, we present the results of our systematic review. The search queries defined in Section \ref{subsec:search_strategy} resulted in the total of 1395 potentially relevant publications:  678 papers from IEEE Xplore, 34 papers from SpringerLink, and 683 papers from ACM Digital Library.

\begin{figure}[h]
    \centering
    \includegraphics[width=0.4\textwidth]{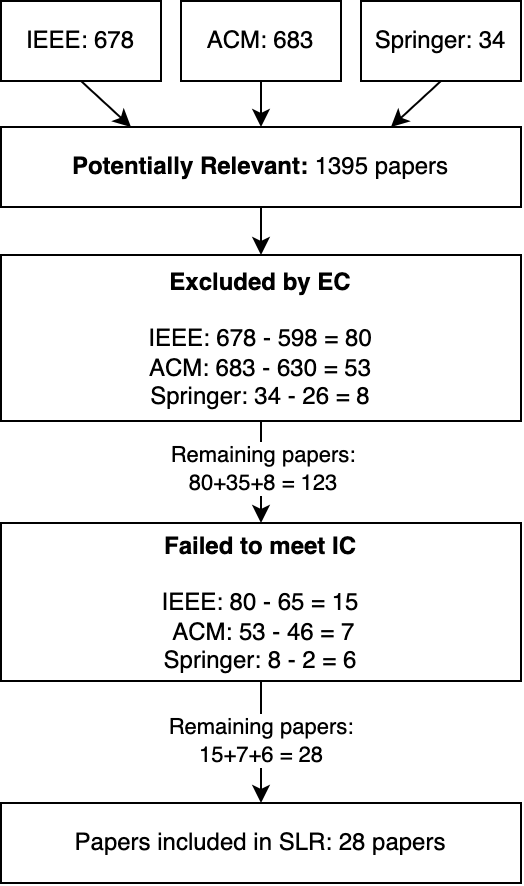}
    \captionsetup{justification=centering}
    \caption{Selection flow chart diagram}
    \label{selection_flow}
\end{figure}

The initial set of 1395 papers were first filtered through the exclusion criteria, and then checked against the inclusion criteria to identify the works most relevant for our SLR (Fig \ref{selection_flow}). After filtering the papers through exclusion and inclusion criteria, we identified 28 papers for further analysis in accordance with criteria defined in Section \ref{subsec:analysis_criteria}.

\begin{figure}[h]
    \centering
    \includegraphics[width=0.45\textwidth]{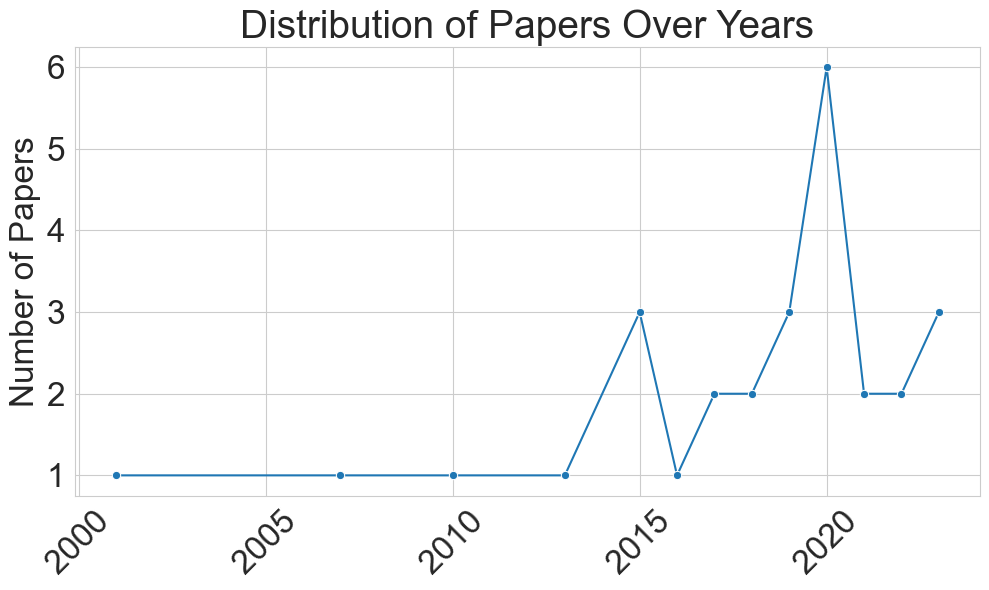}
    \caption{Distribution of papers over years}
    \label{fig:papers_per_year}
\end{figure}

First, we clustered the selected papers based on the publication year and the scientific database. Fig \ref{fig:papers_per_year} shows the distribution of the number of papers published each year. We see that research interest in the area of study has been steady since 2015, peaking in 2020.

\begin{figure}[h]
    \centering
    \includegraphics[width=0.45\textwidth]{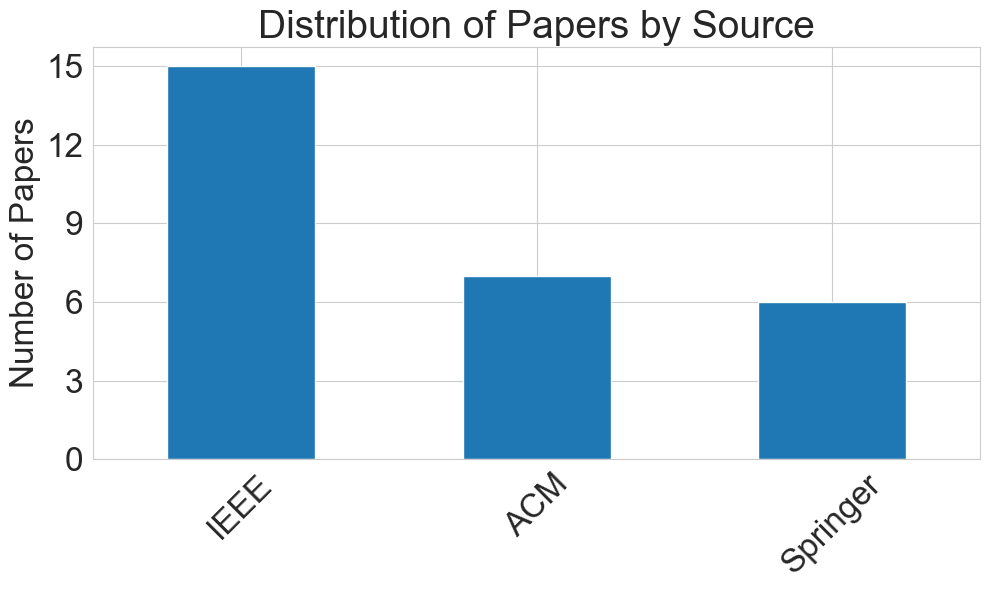}
    \caption{Distribution of the selected papers by database}
    \label{fig:papers_per_source}
\end{figure}

Fig \ref{fig:papers_per_source} shows the the distribution of papers over source databases. IEEE has the highest count, possibly due to having several journals dedicated to educational technologies, while Springer and ACM have approximately the same number of relevant publications. We also conducted a subjective quality evaluation for each of the selected papers, which can be found in Appendix \ref{appendix:quality_scores}. For the reader's convenience, we also present a descriptive summary of each paper in Appendix \ref{appendix:summaries}.

In the following subsections, we first present the general overview of the selected papers, and then proceed to the key results related to the research questions.

\subsection{General overview: Educational context, supported tasks, and recommendation methods}
\label{subsec:general_overview}

We identified three major educational contexts in which the systems were deployed. For a few papers, we were not able to pinpoint a specific educational context, as the description was general and applicable to any of the three context. As a result we had these four context categories.
\begin{itemize}
    \item Formal Education: Deployed within universities, schools, and online courses, focusing on curriculum-based learning.
    \item Informal Learning: Used in non-formal education settings, such as online forums, social media platforms, or MOOCs, for self-directed learning.
    \item Lifelong Learning: Recommending resources for continuous professional development and personal interest learning outside traditional educational structures.
    \item Non-specified: Recommenders that did not mention any particular educational setting.
\end{itemize}

\begin{figure}[h]
    \centering
    \includegraphics[width=0.35\textwidth]{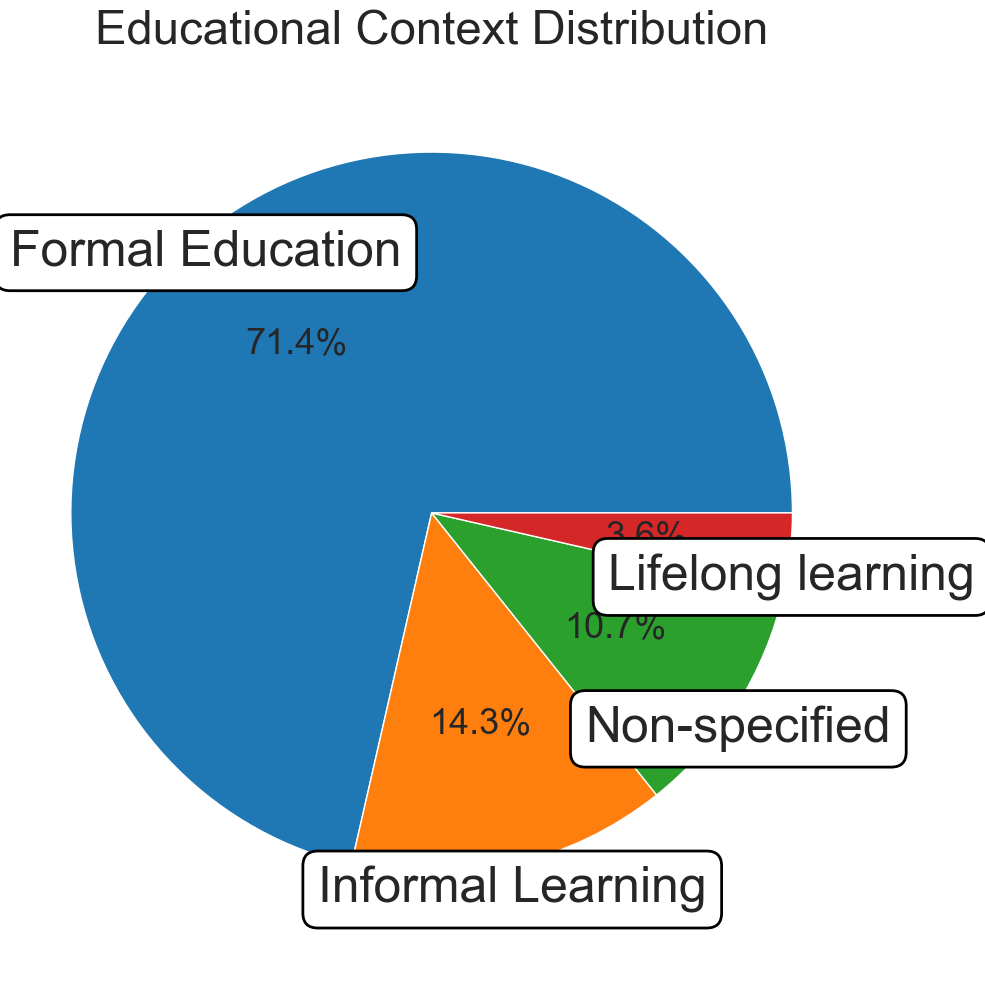}
    \caption{Educational Context Distribution}
    \label{fig:context_distribution}
\end{figure}

Fig \ref{fig:context_distribution} shows the educational contexts' distribution. By far, the most common context in which the recommender systems were deployed was formal education. Among them were higher education \cite{exercise_generation_ontology, pedagogical_framework, discussion_fora, blended7104183, coding_literacy, MLAcademia8693719, largescale, PariserumPerumal2019, scratch8651403, RecommendActions9249379, SubjectEnrollment9226409, AR_DL_computational_thinking, userPreferencesUsingHybridOptimization, ersdoDynamicOntology, theatre_context, recommend_review_materials}, school education \cite{game_vs_its, usercentric7994718}, children education \cite{children9956839}, and distance learning \cite{theatre_context}.

There were four papers in the informal learning domain, three of them \cite{usageContextBased6980102, emotion9434237, ontology10} were a web-based e-learning application, and one social learning platform \cite{SocialRetrieval2018}. Lastly, three studies did not specify any particular context \cite{fuzzyTree7094243, contexual_bandits, heterogenous_evolution_network}. In principle, these three can be deployed in any context as long as it is modelled according to the recommender's requirements. The only paper from the lifelong learning domain was about professional training \cite{Semantic7272748}.

\begin{figure}[h]
    \centering
    \includegraphics[width=0.32\textwidth]{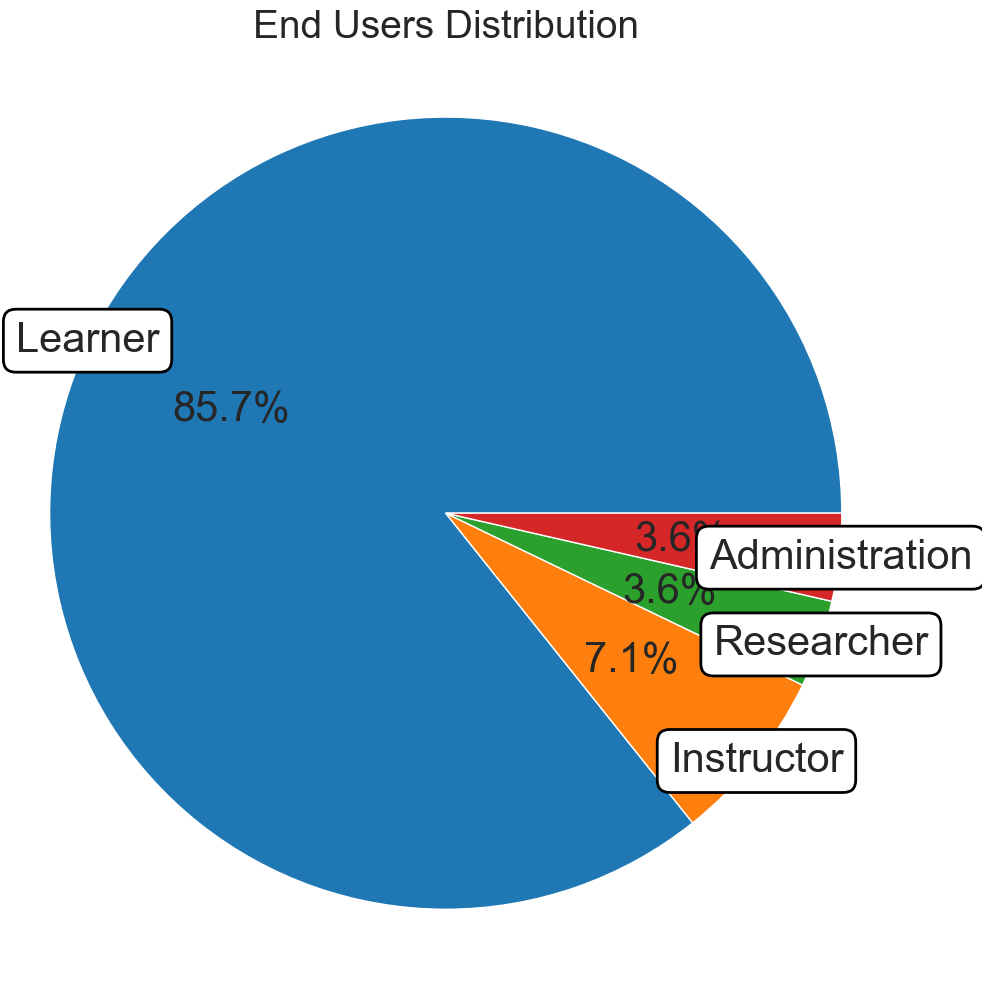}
    \caption{Recommendation end users}
    \label{fig:end_user}
\end{figure}

24 out of 28 papers had learners as the end-users receiving recommendations (Fig. \ref{fig:end_user}). Apart from that, two papers were meant for instructors, one for researchers, and one for university administration. More specifically, \cite{usercentric7994718} recommended resources and online communities to school teachers, \cite{RecommendActions9249379} suggested course improvement actions to university instructors,  \cite{SocialRetrieval2018} suggested research collaborators based on social profile data, and \cite{MLAcademia8693719} automated course allocation and supervisor assignment for university administration.

\begin{figure}[h]
    \centering
    \includegraphics[width=0.45\textwidth]{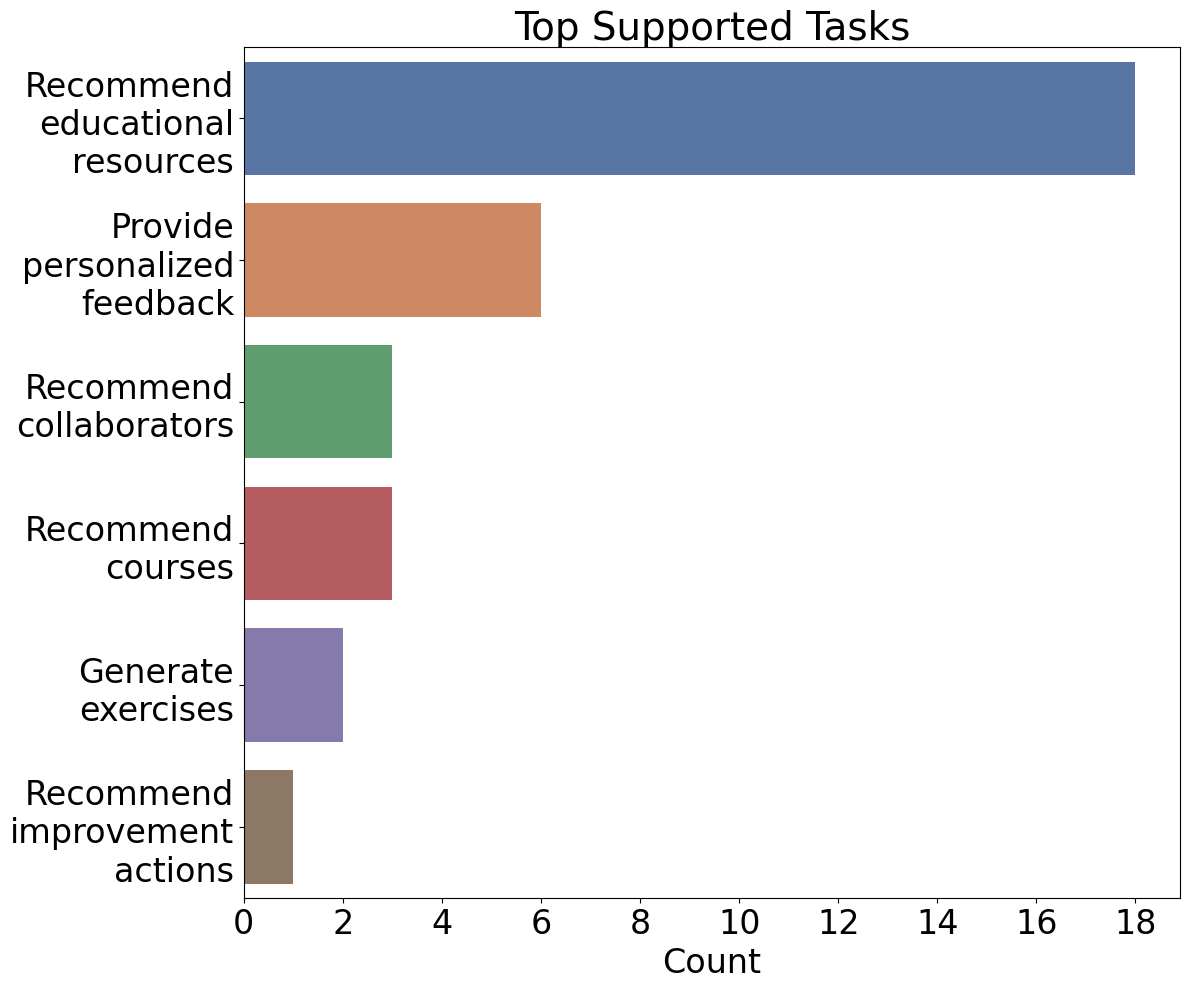}
    \caption{Supported Tasks Distribution}
    \label{fig:supported_tasks}
\end{figure}

As for the supported tasks (Fig. \ref{fig:supported_tasks}), the majority of the RS were designed to recommend educational resources \cite{discussion_fora, fuzzyTree7094243, usageContextBased6980102, Semantic7272748, blended7104183, game_vs_its, usercentric7994718, SocialRetrieval2018, PariserumPerumal2019, scratch8651403, emotion9434237, AR_DL_computational_thinking, contexual_bandits, ontology10, userPreferencesUsingHybridOptimization, children9956839, heterogenous_evolution_network, recommend_review_materials}. These resources can encompass a variety of materials and activities to support learning, from articles and videos to quizzes and interactive exercises. A similar, but less common task was to generate personalized exercises \cite{exercise_generation_ontology, coding_literacy}. The second most popular task was providing personalized feedback \cite{pedagogical_framework, pomdp_representations, blended7104183, game_vs_its, AR_DL_computational_thinking, theatre_context}, which is a common feature for Intelligent Tutoring Systems (ITS). Three systems recommended courses \cite{SubjectEnrollment9226409, largescale, ersdoDynamicOntology} to help learners navigate their way through a plethora of available options based on their academic profile and goals. Three more systems \cite{blended7104183, SocialRetrieval2018, MLAcademia8693719} recommended research collaborators and supervisors. One unique task is the recommendation of course improvement actions to instructors \cite{RecommendActions9249379}, showing that RS can also be tailored to enhance teaching practices. Four of the systems mentioned above \cite{blended7104183, game_vs_its, SocialRetrieval2018, AR_DL_computational_thinking} supported more than one task.

\begin{figure}[h]
    \centering
    \includegraphics[width=0.45\textwidth]{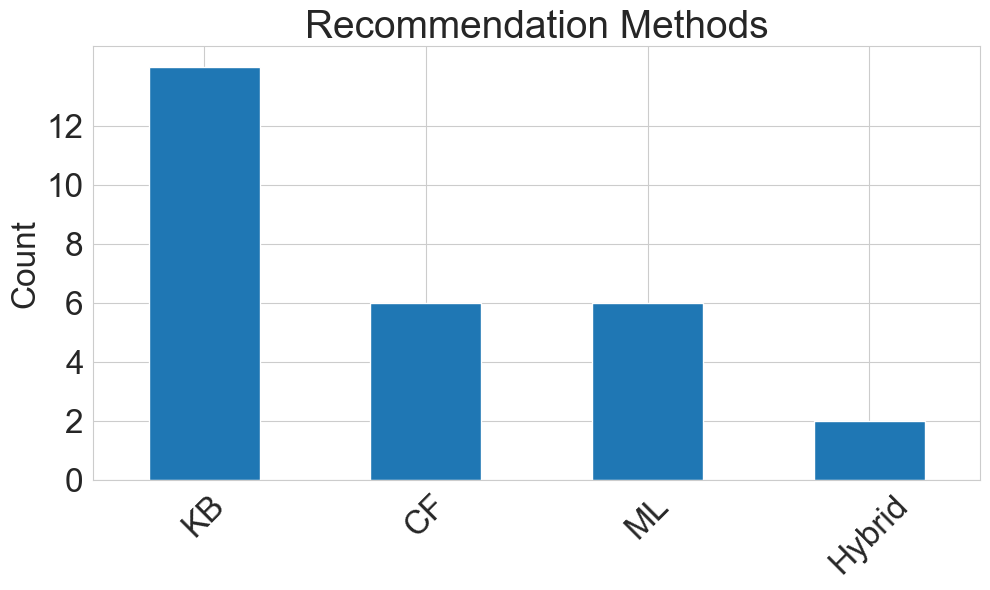}
    \caption{Recommendation Methods: Knowledge-based (KB), Collaborative filtering (CF), Machine-Learning based (ML), and Hybrid.}
    \label{fig:recommendation_method}
\end{figure}

Fig. \ref{fig:recommendation_method} shows recommendation methods used. Knowledge-based (KB) approach is the most prevalent approach used in 14 papers \cite{exercise_generation_ontology, pedagogical_framework, pomdp_representations, usageContextBased6980102, Semantic7272748, blended7104183, game_vs_its, SocialRetrieval2018, MLAcademia8693719, PariserumPerumal2019, ontology10, children9956839, theatre_context, recommend_review_materials}. By definition, this method requires some insight into the relations between the recommended items and learners to provide recommendations. Collaborative Filtering (CF), a method that suggests items based on user ratings, was used in six papers \cite{discussion_fora, usercentric7994718, largescale, scratch8651403, emotion9434237, userPreferencesUsingHybridOptimization}. Machine Learning-based (ML) recommendation methods, such as classification algorithms and predictive modeling, were also utilized in six studies \cite{coding_literacy, RecommendActions9249379, SubjectEnrollment9226409, AR_DL_computational_thinking, contexual_bandits, heterogenous_evolution_network}.
Lastly, two papers employ a hybrid approach that combines different recommendation strategies \cite{fuzzyTree7094243,ersdoDynamicOntology}. Interestingly, no papers used Content-based filtering, possibly due to the fact that the knowledge-based and hybrid approaches serve as a more sophisticated replacement, especially when using ontologies to model the domain items \cite{ersdoDynamicOntology,Semantic7272748,ontology10, theatre_context}.

To summarize the general overview, the main context is formal education, the end-user is mainly learner, and most popular supported task is recommending educational resources. There were also a few papers in the informal learning domain, while the second most popular supported task was providing personalized feedback such as hints. Among the recommendation methods, knowledge-based approach is the most popular, followed by collaborative filtering and machine learning.

In the next subsection, we review the key analysis results directly related the research questions.

\subsection{Key results: Analysing the Research Questions}
\label{subsec:key_results}

In this subsection, we review the analysis criteria directly relating to the research questions (section \ref{subsec:rqs}). The analysis data is shown in Table \ref{tab:main_data}.

\begin{table}[!htbp]
    \centering
    \caption{Main results based on the analysis criteria}
    \label{tab:main_data}
    \begin{tabularx}{0.49\textwidth}{|p{0.5cm}|X|p{8mm}|p{21mm}|p{8mm}|}
        \hline
        \textbf{Ref} & \textbf{Target Metric} & \textbf{Includes LOs} & \textbf{Evaluation Method} & \textbf{Assesses LOs} \\
        \hline \cite{AR_DL_computational_thinking} & \parbox[t][0.6cm]{3cm}{Academic Performance \\
        Self-Evaluation} & Yes & \parbox[t][0.6cm]{3cm}{Formal Assessment \\
        Qualitative} & Yes \\
        \hline \cite{recommend_review_materials} & \parbox[t][0.9cm]{3cm}{Academic Performance \\
        User Rating} & Yes & \parbox[t][0.9cm]{3cm}{Formal Assessment \\
        Experimental \\
        Qualitative} & Yes \\
        \hline \cite{game_vs_its} & \parbox[t][0.6cm]{3cm}{Academic Performance \\
        User Satisfaction \\
        Self-Evaluation} & Yes & \parbox[t][0.9cm]{3cm}{Formal Assessment \\
        Experimental \\
        Qualitative} & Yes \\
        \hline \cite{scratch8651403} & \parbox[t][0.9cm]{3cm}{Academic Performance \\
        User Satisfaction} & Yes & \parbox[t][0.9cm]{3cm}{Formal Assessment \\
        Experimental \\
        Qualitative} & Yes \\
        \hline \cite{blended7104183} & \parbox[t][0.6cm]{3cm}{Academic Performance} & Yes & \parbox[t][0.6cm]{3cm}{Formal Assessment \\
        Experimental} & Yes \\
        \hline \cite{pomdp_representations} & \parbox{3cm}{Probability of Learning} & Yes & \parbox{3cm}{Simulation} & Yes \\
        \hline \cite{children9956839} & \parbox{3cm}{Probability of Learning} & Yes & \parbox{3cm}{Simulation} & Yes \\
        \hline \cite{pedagogical_framework} & \parbox{3cm}{Probability of Learning} & Yes & \parbox{3cm}{No evaluation} & Yes \\
        \hline \cite{coding_literacy} & \parbox{3cm}{Probability of Learning} & Yes & \parbox{3cm}{No evaluation} & Yes \\
        \hline \cite{heterogenous_evolution_network} & \parbox{3cm}{Academic Performance} & Yes & \parbox{3cm}{Benchmarking} & No \\
        \hline \cite{RecommendActions9249379} & \parbox{3cm}{Academic Performance} & Yes & \parbox{3cm}{Experimental} & No \\
        \hline \cite{Semantic7272748} & \parbox{3cm}{Expert rating} & Yes & \parbox{3cm}{Experimental} & No \\
        \hline \cite{fuzzyTree7094243} & \parbox{3cm}{User Rating} & Yes & \parbox{3cm}{Benchmarking} & No \\
        \hline \cite{SubjectEnrollment9226409} & \parbox{3cm}{Academic Performance} & No & \parbox{3cm}{Experimental} & No \\
        \hline \cite{theatre_context} & \parbox{3cm}{Expert rating} & No & \parbox{3cm}{Experimental} & No \\
        \hline \cite{emotion9434237} & \parbox[t][0.6cm]{3cm}{User Rating \\
        User Satisfaction} & No & \parbox[t][0.6cm]{3cm}{Experimental \\
        Qualitative} & No \\
        \hline \cite{usageContextBased6980102} & \parbox{3cm}{User Rating} & No & \parbox{3cm}{Benchmarking} & No \\
        \hline \cite{contexual_bandits} & \parbox{3cm}{User Rating} & No & \parbox{3cm}{Benchmarking} & No \\
        \hline \cite{ersdoDynamicOntology} & \parbox{3cm}{User Rating} & No & \parbox{3cm}{Benchmarking} & No \\
        \hline \cite{discussion_fora} & \parbox{3cm}{User Rating} & No & \parbox{3cm}{Experimental} & No \\
        \hline \cite{MLAcademia8693719} & \parbox{3cm}{User Rating} & No & \parbox{3cm}{Experimental} & No \\
        \hline \cite{PariserumPerumal2019} & \parbox{3cm}{User Rating} & No & \parbox{3cm}{Experimental} & No \\
        \hline \cite{largescale} & \parbox{3cm}{User Rating} & No & \parbox{3cm}{No evaluation} & No \\
        \hline \cite{SocialRetrieval2018} & \parbox{3cm}{User Rating} & No & \parbox{3cm}{Qualitative} & No \\
        \hline \cite{ontology10} & \parbox{3cm}{User Rating} & No & \parbox{3cm}{Qualitative} & No \\
        \hline \cite{userPreferencesUsingHybridOptimization} & \parbox{3cm}{User Rating} & No & \parbox{3cm}{Qualitative} & No \\
        \hline \cite{exercise_generation_ontology} & \parbox{3cm}{User Satisfaction} & No & \parbox{3cm}{Qualitative} & No \\
        \hline \cite{usercentric7994718} & \parbox{3cm}{User Satisfaction} & No & \parbox{3cm}{Qualitative} & No \\
        \hline
    \end{tabularx}
    \label{tab:specific_analysis_results}
\end{table}

For a more detailed account of results, please see our Github repository with the CSV file containing all collected review data \footnote{\url{https://github.com/nurlingo/slr-ers/blob/main/data.csv}}.\\
\\
\emph{RQ1: What target metrics are optimized in educational recommender systems (ERS)?}

To answer \emph{RQ1}, we examined Target Metrics (\emph{AC1}) that each of the examined recommenders aimed to increase.

\begin{figure}[h]
    \centering
    \includegraphics[width=0.45\textwidth]{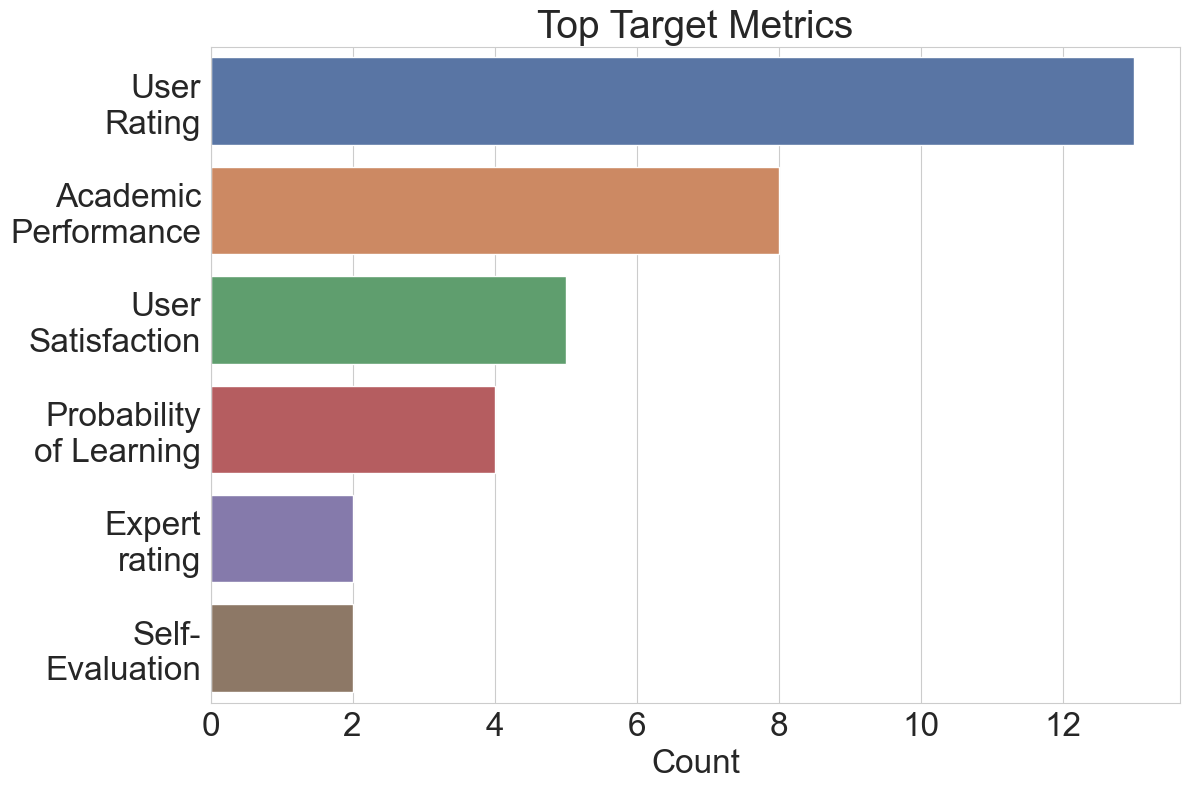}
    \caption{AC1: Target Metrics that ERS optimize}
    \label{fig:target_metrics}
\end{figure}

The most popular target metric is user-rating. Almost half of the papers used it when evaluating the effectiveness of their system \cite{discussion_fora, fuzzyTree7094243, usageContextBased6980102, SocialRetrieval2018, MLAcademia8693719, largescale, PariserumPerumal2019, emotion9434237, contexual_bandits, ontology10, userPreferencesUsingHybridOptimization, ersdoDynamicOntology}. The user ratings were collected either explicitly by allowing users to rate the items or implicitly by logging the desired user behaviours such as clicks. Two more papers \cite{theatre_context, Semantic7272748} relied on expert opinion and compared their system's performance against recommendations given by experts. Importantly, both user and expert ratings were in fact proxies for relevance to the user preferences and needs.

A similar user-centered set of target metrics were user-satisfaction \cite{game_vs_its, usercentric7994718, scratch8651403, emotion9434237} and self-evaluation \cite{game_vs_its, AR_DL_computational_thinking}. These are similar to user rating-based relevance, however present a more diverse set of criteria to evaluate the recommendations'. The users were asked about the effect on their learning and their satisfaction. Interestingly, \cite{game_vs_its} demonstrates that higher satisfaction and self-evaluation do not necessarily correlate with academic performance.

Twelve papers used more learning-centered metrics. The most popular among them one was academic performance \cite{blended7104183, game_vs_its, scratch8651403, RecommendActions9249379, SubjectEnrollment9226409, AR_DL_computational_thinking, heterogenous_evolution_network, recommend_review_materials}, mostly implemented within the context of higher education and measured through formal course assessment.

Another learning-centered metric was probability of learning that assesses how likely it is that a learner has mastered a concept or a skill based on observing the learner behaviour. Three papers \cite{pedagogical_framework, pomdp_representations, coding_literacy} implemented it in the context of an intelligent tutoring or adaptive learning system, where they modelled the learner through knowledge states and observed his performance to update the state probabilities. While \cite{children9956839} used the number of interactions with other learners to determine the likelihood of learning.

Worth noting that several papers \cite{game_vs_its, scratch8651403, emotion9434237, AR_DL_computational_thinking, recommend_review_materials} used two or more target metrics, thus allowing for a more comprehensive view that can cover both user-centered and learning-centered perspectives.\\
\\
\emph{RQ2: What evaluation methods are used for measuring the effectiveness of ERS?}

To answer RQ2, we reviewed the evaluation methods (\emph{AC2}) in selected papers, and placed them into a few broad categories.

\begin{figure}[h]
    \centering
    \includegraphics[width=0.45\textwidth]{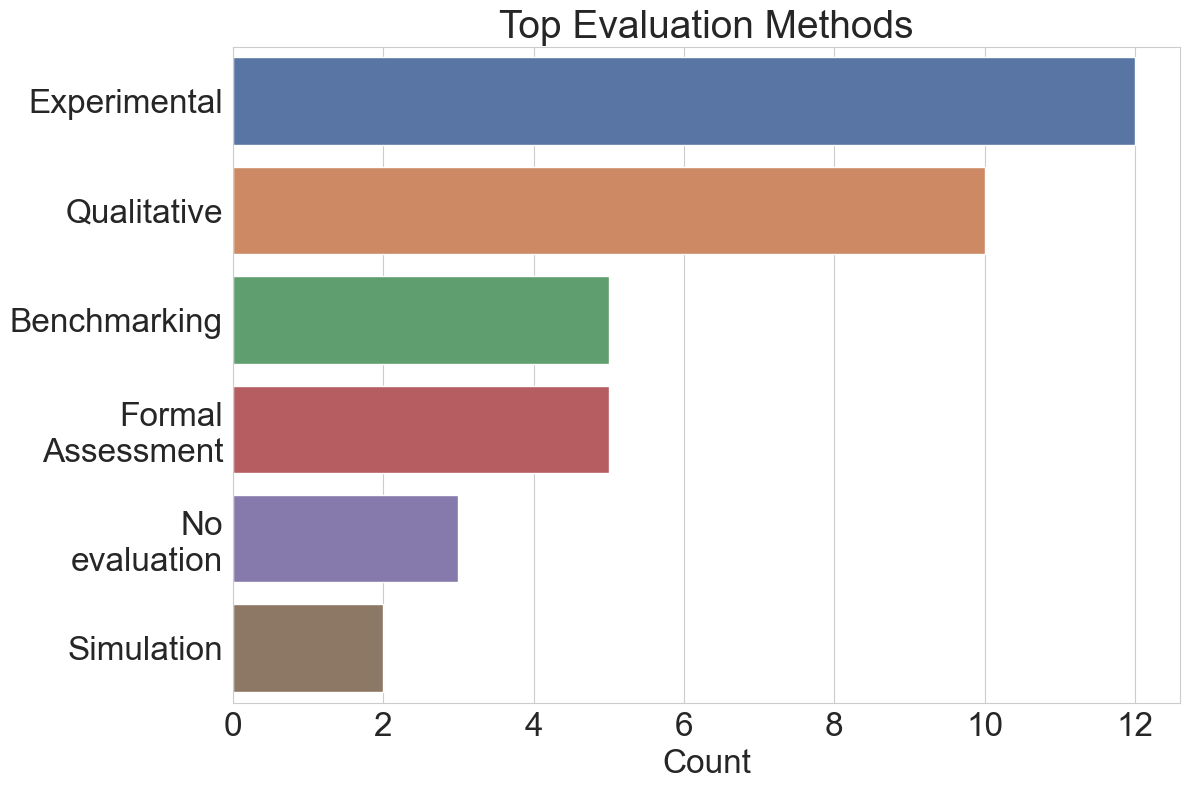}
    \caption{AC2: Evaluation methods used}
    \label{fig:evaluation_methods}
\end{figure}

A significant portion of the studies \cite{discussion_fora, Semantic7272748, blended7104183, game_vs_its, MLAcademia8693719, PariserumPerumal2019, RecommendActions9249379, SubjectEnrollment9226409, scratch8651403, emotion9434237, recommend_review_materials, theatre_context} have opted for experimental evaluation methods including online controlled experiments with pre- and post-tests, as well as offline data splits into training and test datasets.

Ten studies \cite{exercise_generation_ontology, game_vs_its, SocialRetrieval2018, usercentric7994718, AR_DL_computational_thinking, scratch8651403, emotion9434237, ontology10, userPreferencesUsingHybridOptimization, recommend_review_materials} have utilized qualitative measurements questionnaires to understand the user experience and satisfaction with ERS.

Five studies \cite{blended7104183, game_vs_its, AR_DL_computational_thinking, scratch8651403, recommend_review_materials} used formal academic assessment, mostly in combination with some experimental setting, whereby learners were split into groups and their performance was assessed through course work, tests, and quizzes. Five more studies \cite{fuzzyTree7094243, usageContextBased6980102, contexual_bandits, ersdoDynamicOntology, heterogenous_evolution_network} employed benchmarking approaches to evaluate their systems against other known algorithms using public datasets. Two studies \cite{pomdp_representations, children9956839} used simulation (a virtual model of the system) and studied its behavior under various conditions. While a few studies had no relevant evaluation stage \cite{pedagogical_framework, coding_literacy, largescale}.

Some researchers used mixed methods to provide a more comprehensive evaluation of their ERS. For instance, \cite{game_vs_its, AR_DL_computational_thinking, emotion9434237, usercentric7994718, SocialRetrieval2018} combined experimental methods and qualitative metrics to bridge the gap between objective measurements and subjective user experiences.

\begin{figure}[h]
    \centering
    \includegraphics[width=0.45\textwidth]{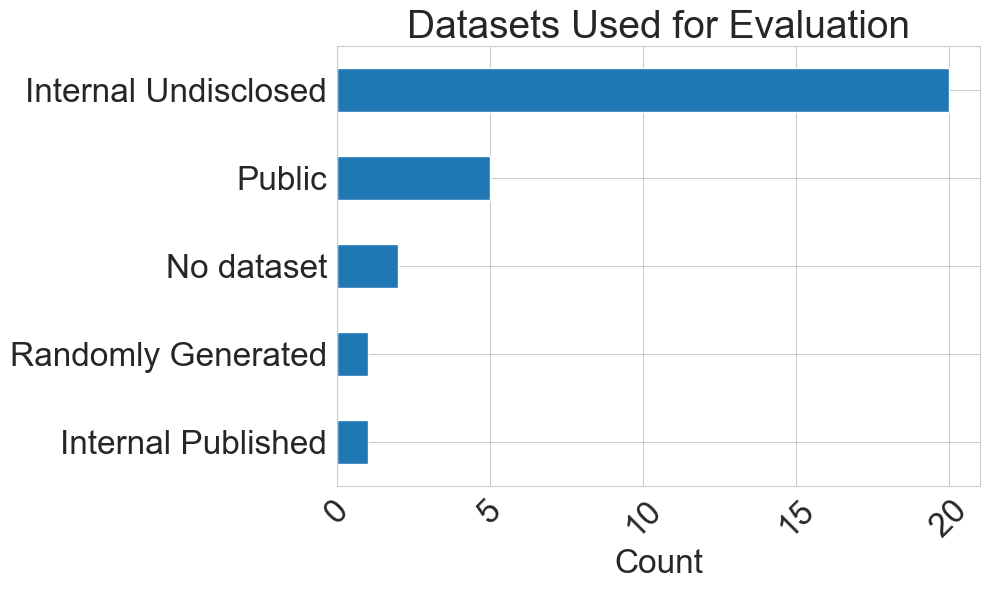}
    \caption{Evaluation datasets used in the papers}
    \label{fig:datasets}
\end{figure}

As a side note, we noticed that the majority of studies relied on undisclosed, internal datasets (Fig. \ref{fig:datasets}) when conducting evaluation. One of the studies used famous MovieLens \cite{fuzzyTree7094243}, which is a dataset from an unrelated domain. This does not allow for the reproducibility of the studies, and possibly indicates a lack of available educational datasets.

\emph{RQ3: How do educational recommender systems implement outcome-based assessment?}

To answer \emph{RQ3}, we reviewed whether the systems used learning outcomes (\emph{AC3}) and outcome-based assessment (\emph{AC4}) within the recommendation process.

\begin{figure}[h]
    \centering
    \includegraphics[width=0.45\textwidth]{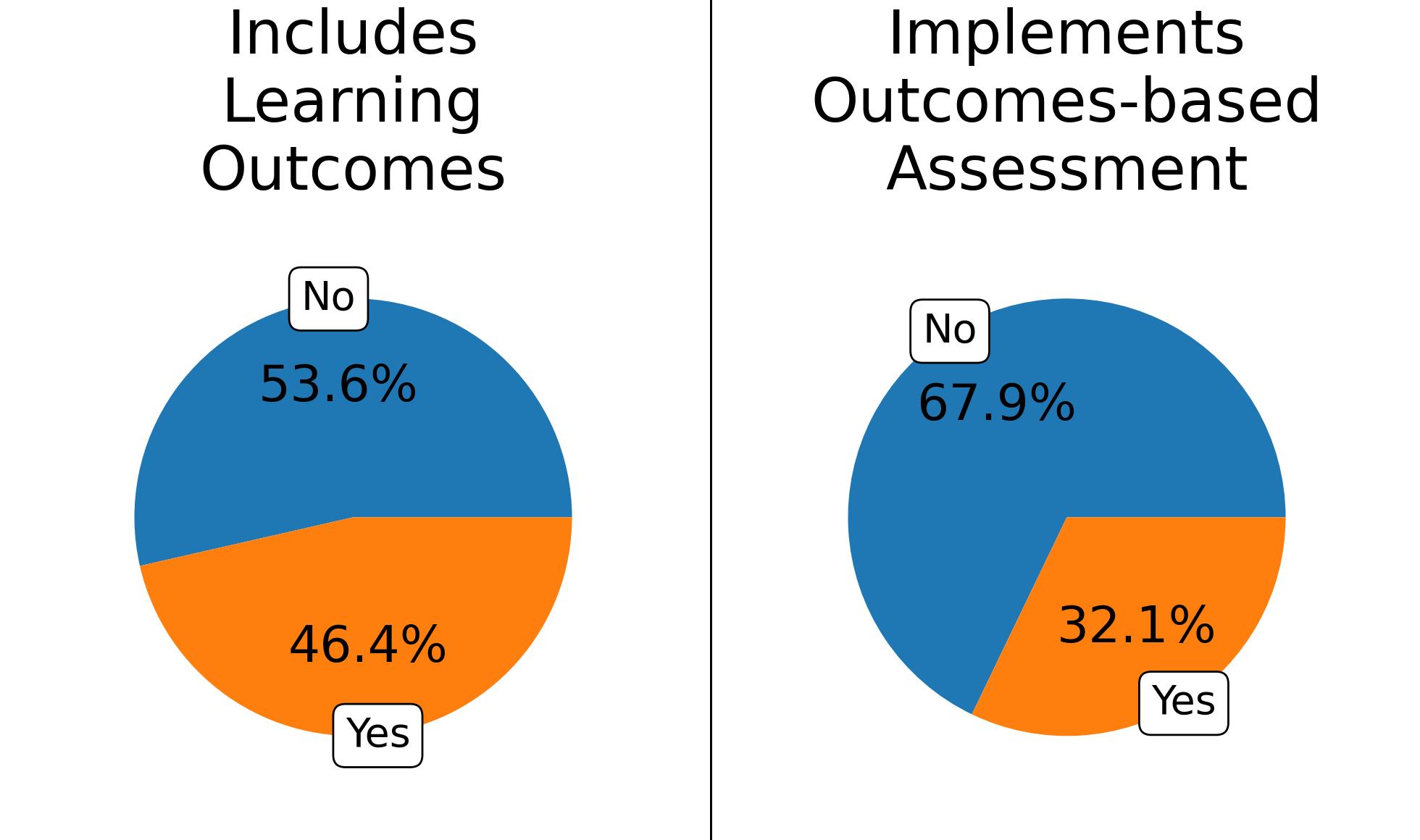}
    \caption{Inclusion and Assessment of Learning Outcomes}
    \label{fig:learning_outcomes}
\end{figure}

As shown in Fig. \ref{fig:learning_outcomes}, almost half of the papers made use of learning outcomes. Six of these papers \cite{usercentric7994718, blended7104183, RecommendActions9249379, AR_DL_computational_thinking, recommend_review_materials, scratch8651403} recommended items within a university course, and therefore reused the intended outcomes readily available for the course.

Five other papers \cite{heterogenous_evolution_network, game_vs_its, pedagogical_framework, pomdp_representations, coding_literacy, children9956839} defined outcomes manually as concepts or skills relevant to the subject being studied and modelled them as part of the recommendation process. A similar approach was taken by \cite{Semantic7272748, fuzzyTree7094243}, who defined outcomes in the form of professional competencies and used them as the basis for making recommendations. The rest of the papers either did not include learning outcomes or mode no explicit mention of their use.

As for the outcome-based assessment, roughly a third of papers measured the achievement of learning outcomes (Fig. \ref{fig:learning_outcomes}). Seven of them \cite{fuzzyTree7094243, blended7104183, AR_DL_computational_thinking, recommend_review_materials, scratch8651403, game_vs_its, pomdp_representations} used formal assessment readily available as part of a university course or professional training. In that case, the learners were split into several group for the evaluation through a controlled experiment.

Three more papers could be said to have automated outcome-based assessment. \cite{pedagogical_framework, coding_literacy} assessed the solution to automatically generated programming exercises. In their case, different parts of the expected solution were linked to a particular learning outcome, and the probability of achieving the outcome was updated based on observing which parts of the program were written correctly. While \cite{children9956839} took an indirect approach and approximated the assessment by putting a threshold on how many activations (engagements with other learners) should a learner have to be marked as someone who learnt a particular concept.

Interestingly, four papers included learning outcomes, but did not conduct the assessment. \cite{heterogenous_evolution_network, RecommendActions9249379} recommended items that had the highest possibility of success based on historic data, but did not verify their predictions through assessment. While \cite{Semantic7272748, fuzzyTree7094243} prioritized relevance-based target metrics.

To summarize, our objective was to review how educational recommender systems (ERS) optimize learning.  We reviewed the target metrics used in existing ERS with the corresponding evaluation methods. Since we define learning as the achievement of learning outcomes, we also reviewed how they implemented outcome-based assessment. The summary results of our review are presented in Table \ref{tab:main_data}. In the following section, we elaborate our findings in the light of research questions, offer interpretations, and discuss some of the limitations and further work.

\section{Discussion}
\label{sec:discussion}

Our SLR on Educational Recommender Systems (ERS) reveals several important trends, as well as some potential gaps that merit further exploration. In the general overview, we found that ERS were mostly deployed in the context of formal education, the end-users were mostly learners, and the recommended items were mainly educational resources. The most common approach was knowledge-based, which shows that researchers try to apply their understanding of the domain to improve the recommendations. Since these overview results (Section \ref{subsec:general_overview}) have been covered by and generally agree with previous systematic reviews on ERS \cite{urdaneta2021recommendation, da2023systematic}, we immediately proceed to discuss the key results related to the research questions of our SLR.

\subsection{Lack of focus on learning}

For the target metric, most of the reviewed ERS optimized relevance of the served items, then learning, then other user-centric metrics such as satisfaction. This raises a question: why relevance is the most popular target metric in the domain of education, and not a metric directly related to learning? Given that the learner is the most common end-user and educational resources are the main focus of recommendations, one might assume that learning should be primary focus, yet only 12 papers out of 28 optimized learning.

One possible explanation is the historical context within which recommender systems emerged. The focus on the relevance and classification or prediction performance metrics aligns with the needs in entertainment and e-commerce, where the recommended item is closer to the business goal. Possibly, the researchers transferred the algorithms together with evaluation methods without necessary adjustments to the domain. However, even in these initial domains, they have come to use more sophisticated long-term metrics such as retention and promoting habitual use \cite{lukyanchikova2023case}. This means that educational recommenders should also adjust the evaluation metrics, i.e. move beyond relevance and closer to the business goals within the domain. This highlights the need for determining a comprehensive overall evaluation criteria (OEC) \cite{fabijan2017evolution} that includes educational, user-centric, and product-oriented metrics. Interestingly, we can already see that some of the reviewed papers used a combination of several target metrics to either compare between them or to have a more comprehensive view. This also correlates with the recent trend of having context-rich user profiles in ERS \cite{review_context6189308}, whereby researchers are exploring new significant user properties to base the recommendations on. As further work, we can combine the outcome-based approach with context-rich recommendations to determine the best predictors. This would answer the question of where should the attention go when recommending the items and assessing their effectiveness from the pedagogical point of view.

Another possible explanation is that the researchers operate on the assumption that relevance leads to engagement, while engagement leads to greater learning outcomes. This is partially supported by previous research \cite{chi2014icap}. However, this approach does not allow to differentiate between more and less effective learning activities, since it assumes the activity to be effective as long as the learner is engaged. Several of the reviewed papers proposed more holistic user-centric metrics to ensure the recommendations are having an overall positive effect. However these are still not outcome-focused, as both engagement, satisfaction, and positive self-evaluation has been shown to not necessarily correlate with learning \cite{game_vs_its}. Moreover, using indirect metrics to measure learning presents a limitation from the scientific point of view, since having indirect observations will necessarily be less conclusive. Evaluating the learning through more direct means would enable us to determine the most effective types of engagement, context elements, and recommendation methods from the pedagogical point of view. This also agrees with the principal idea in engineering that metrics should correspond to business goals \cite{gqm}, and the similar idea in pedagogy that learning activities and assessment should be aligned with the intended outcomes \cite{biggs2012student}.

If standard engagement metrics are used, they should be adjusted to focus more on educational relevance than on general popularity or click-through rate. For example, engagement metrics can include time spent on learning tasks, completion rates of recommended courses or activities, and interaction rates with the system, hence ultimately contributing to and correlating with the achievement of learning outcomes. Retention can be reformulated as progression or completion rates to track the learners' progress in a course or program over time. Similarly user-centric metrics can be important in ERS, if the satisfaction is closely tied to educational value and personal learning progress. Personalization would then mean the alignment of recommendations with individual learning goals, styles, or other contextual learner properties. However, none of the examined papers frame the discussion of metrics in this manner. An important direction for further work is to investigate the correlation between learning, engagement, and satisfaction, and thus attemp to identify the types of indirect metrics most correlating with learning to optimize towards them.

Yet another explanation is that learning is hard to measure \cite{learning_measurement, assessment_challenges}. In other words, despite understanding the importance of assessing the learning and using it as the principle optimization target, the researchers are bound to use other means. It can be argued that learning assessment is a wicked problem \cite{zhao2019tackling}, which is hard to define and measure definitively. It can also be argued that the outcome-oriented approach is not always applicable, for example in affective learning domain \cite{picard2004affective}, where learning outcomes are open-ended rather than pre-defined by the instructor. However, open-endedness does not mean absence, and even in the affective learning the best practice is to provide a way for the learner to self-determine, document, and reflect upon the achieved learning outcomes.

Being the proponents of the outcome-based approach, we recognize that designing well-aligned learning assessment is a demanding and resource-intensive task. It requires domain knowledge, understanding the learner's level and needs, determining the intended outcomes, and designing the assessment activities. This brings us to \emph{RQ2} and \emph{RQ3} which explore the evaluation methods used by ERS and how outcome-based assessment is implemented.

\subsection{The relation between target metrics and evaluation methods}
% RQ2
The reviewed research had several evaluation methods evenly distributed over the papers. The menthods included experimental, qualitative, simulation, benchmarking, and mixed approaches. This distribution reflects a broad consensus on the importance of multifaceted evaluation.

Our main interest was in the methods used with the learning-centered target metrics. Those were mostly gathered through formal assessment combined with experimental methods. Formal assessment is possible in formal education, where the learning outcomes and corresponding assessment activities are defined at the outset. However, this can be challenging in informal learning domain due to a lack of structured assessment mechanisms. Two of the reviewed works attempted to assess the probability of learning through simulation, and two more works omitted evaluation of their proposed system despite having a learning-centric target metric. This supports the explanation above that learning is hard to measure, especially so in the informal domain, as the ERS that evaluated learning effectiveness mostly did so in the context of academic courses.

An interesting question is how much the feasibility of an evaluation method determines the target metric. In the ideal world, the metrics should proceed from goals, and then the corresponding method should be chosen. Possibly, there are not many public datasets sufficiently rich to allow a comprehensive evaluation of recommendations from the learning point of view. Most of the open educational datasets used in the papers \cite{usercentric7994718, ersdoDynamicOntology, usageContextBased6980102, MLAcademia8693719} focus on engagement and ratings data. Therefore, the absence of appropriate datasets might be a factor affective the choice of target metrics. While having such datasets would enable researchers to conduct more meaningful evaluations of educational recommender systems, and promote the essential objective of education - ensuring the achievement of specific learning outcomes.

% RQ3:
\subsection{Three common ways to implement outcome-based assessment}
For outcome-based assessment, we can essentially outline three approaches that were used within the reviewed ERS: outcomes and assessment were already in place as part of a formal course, or the assessment exercises automatically generated and graded, or their achievement was concluded from indirect observations.

The first approach is when ERS is deployed within a structured academic course or professional training. In this case, both the outcomes and corresponding formal assessment are already defined, and, in principle, need not to be part of the recommendation process, rather it is enough to log the assessment results and optimize the recommendations accordingly. The evaluation then will be done through splitting the learners into experimental groups. This can work given that the course organizers correctly define the outcomes, and align the assessment activities accordingly. However, this approach does not resolve the question of efficiency for informal settings.

The second approach is essentially a knowledge-based method, where the outcomes, assessment activities and relation between them are modelled within the system. This approach necessitates a dynamic learner, where learner completes the assessment exercises, and the mastery of corresponding outcomes is concluded based on the results. However, this approach requires knowledge of the domain and significant manual effort when composing the outcomes, assessment, and their relations. Therefore, the composition of outcomes and assessment exercises, mapping between them should ideally be automated. If the automation is viable, this can be an effective approach even within the informal learning domain at scale. This is a potential area of further work.

The third approach is to replace traditional assessment with other indirect measures indicative of learning, such as collaboration with other users. However, such a replacement has to strongly correlate with learning. Which raises concerns similar to what has been discussed above in regards to using relevance as the proxy for learning. Still, if such correlation is demonstrated in a specific setting, then the indirect measure could possibly be used instead of assessment activities. Possibly assessment initially be used to identify the most important user-context elements, to then predict the learning gains and maximize the effect of recommendation methods from the pedagogical point of view.

In summary:
\begin{itemize}
    \item There appear to be a gap in outcome-based evaluation of educational recommender systems, especially outside of a formal educational setting. The existing ERS tend to optimize standard recommendation metrics such as rating-based relevance, click-Through Rate (CTR), and user-satisfaction;
    \item ERS should use comprehensive evaluation criteria reflecting the system and domain goals. The main focus should be on learning-oriented metrics, while the standard recommender metrics can be reinterpreted to align with learning.
    \item There are three common approaches for outcome-based assessment in ERS, namely formal academic assessment, knowledge-based modelling, and using indirect measures. Each of the approaches has its limitations and requirements for the deployment context. The first two approaches raise the question of efficiency, while the third approach necessitate the identification of strong predictors of learning.
\end{itemize}

Therefore, we propose three significant areas of focus for further work:
\begin{enumerate}
    \item Defining a comprehensive overall evaluation criteria (OEC) that incorporates outcome-based, user-centric, and product-oriented metrics;
    \item Identifying the key predictors of learning, i.e. metrics and user-context properties that strongly correlate with learning when recommending items;
    \item Experimenting with various efficient ways to implement outcome-based assessmetns, possibly through automation or correlated metrics.
\end{enumerate}

\section{Conclusion}
\label{sec:conclusion}
The SLR explores how learning is measured and optimized in educational recommender systems (ERS). In particular, we investigated the target metrics optimized in ERS, the evaluation methods used for measuring their effectiveness, and how outcome-based assessment is implemented in ERS. Our review began with 1395 papers, which were narrowed down to 28 relevant works. We found rating-based relevance to be the prevalent target metric. We also found that only a third of papers implemented outcomes-based assessment, possibly indicating a gap. Further work in ERS should propose comprehensive evaluation criteria that prioritize learning-related metrics, while still covering user-centric and product-oriented perspectives by adjusting them to the domain. Finding efficient ways to conduct outcome-based assessment appears to be another critical aspect to address the gap.

\bibliographystyle{ieeetr}
\bibliography{references}

\begin{thebibliography}{10}

\bibitem{rivera2018recommendation}
A.~C. Rivera, M.~Tapia-Leon, and S.~Lujan-Mora, ``Recommendation systems in education: a systematic mapping study,'' in {\em International Conference on Information Technology \& Systems}, pp.~937--947, Springer, 2018.

\bibitem{drachsler2015panorama}
H.~Drachsler, K.~Verbert, O.~C. Santos, and N.~Manouselis, ``Panorama of recommender systems to support learning,'' in {\em Recommender systems handbook}, pp.~421--451, Springer, 2015.

\bibitem{urdaneta2021recommendation}
M.~C. Urdaneta-Ponte, A.~Mendez-Zorrilla, and I.~Oleagordia-Ruiz, ``Recommendation systems for education: systematic review,'' {\em Electronics}, vol.~10, no.~14, p.~1611, 2021.

\bibitem{bennett2007netflix}
J.~Bennett, S.~Lanning, {\em et~al.}, ``The netflix prize,'' in {\em Proceedings of KDD cup and workshop}, vol.~2007, p.~35, New York, 2007.

\bibitem{knijnenburg2012explaining}
B.~P. Knijnenburg, M.~C. Willemsen, Z.~Gantner, H.~Soncu, and C.~Newell, ``Explaining the user experience of recommender systems,'' {\em User modeling and user-adapted interaction}, vol.~22, pp.~441--504, 2012.

\bibitem{maslowska2022role}
E.~Maslowska, E.~C. Malthouse, and L.~D. Hollebeek, ``The role of recommender systems in fostering consumers' long-term platform engagement,'' {\em Journal of Service Management}, vol.~33, no.~4/5, pp.~721--732, 2022.

\bibitem{xiao2007commerce}
B.~Xiao and I.~Benbasat, ``E-commerce product recommendation agents: Use, characteristics, and impact,'' {\em MIS quarterly}, pp.~137--209, 2007.

\bibitem{garcia2013educational}
S.~Garcia-Martinez and A.~Hamou-Lhadj, ``Educational recommender systems: A pedagogical-focused perspective,'' in {\em Multimedia Services in Intelligent Environments}, pp.~113--124, Springer, 2013.

\bibitem{gqm}
V.~R. B.~G. Caldiera and H.~D. Rombach, ``The goal question metric approach,'' {\em Encyclopedia of software engineering}, pp.~528--532, 1994.

\bibitem{learning_measurement}
B.~Kizlik, ``Measurement, assessment, and evaluation in education,'' {\em Retrieved October}, vol.~10, p.~2015, 2012.

\bibitem{europeanQualification}
E.~Union, ``Europass tools: European qualifications framework.'' \url{https://europa.eu/europass/en/europass-tools/european-qualifications-framework}.
\newblock [Accessed: May 15, 2023].

\bibitem{biggs2012student}
J.~Biggs, ``What the student does: Teaching for enhanced learning,'' {\em Higher education research \& development}, vol.~31, no.~1, pp.~39--55, 2012.

\bibitem{askarbekuly2020combining}
N.~Askarbekuly, A.~Sadovykh, and M.~Mazzara, ``Combining two modelling approaches: Gqm and kaos in an open source project,'' in {\em Open Source Systems: 16th IFIP WG 2.13 International Conference, OSS 2020, Innopolis, Russia, May 12--14, 2020, Proceedings 16}, pp.~106--119, Springer, 2020.

\bibitem{burke2002hybrid}
R.~Burke, ``Hybrid recommender systems: Survey and experiments,'' {\em User modeling and user-adapted interaction}, vol.~12, no.~4, pp.~331--370, 2002.

\bibitem{fernandez2009challenges}
L.~Fernandez-Luque, R.~Karlsen, and L.~K. Vognild, ``Challenges and opportunities of using recommender systems for personalized health education,'' {\em Medical informatics in a united and healthy Europe}, pp.~903--907, 2009.

\bibitem{review_context6189308}
K.~Verbert, N.~Manouselis, X.~Ochoa, M.~Wolpers, H.~Drachsler, I.~Bosnic, and E.~Duval, ``Context-aware recommender systems for learning: A survey and future challenges,'' {\em IEEE Transactions on Learning Technologies}, vol.~5, no.~4, pp.~318--335, 2012.

\bibitem{RAZA201984}
S.~Raza and C.~Ding, ``Progress in context-aware recommender systems — an overview,'' {\em Computer Science Review}, vol.~31, pp.~84--97, 2019.

\bibitem{deeplearning2022}
L.~Salau, M.~Hamada, R.~Prasad, M.~Hassan, A.~Mahendran, and Y.~Watanobe, ``State-of-the-art survey on deep learning-based recommender systems for e-learning,'' {\em Applied Sciences}, vol.~12, no.~23, p.~11996, 2022.

\bibitem{reinforcement2022}
M.~M. Afsar, T.~Crump, and B.~Far, ``Reinforcement learning based recommender systems: A survey,'' {\em ACM Computing Surveys}, vol.~55, no.~7, pp.~1--38, 2022.

\bibitem{dwivedi2017recommender}
S.~Dwivedi and V.~K. Roshni, ``Recommender system for big data in education,'' in {\em 2017 5th National Conference on E-Learning \& E-Learning Technologies (ELELTECH)}, pp.~1--4, IEEE, 2017.

\bibitem{review_styles10022322}
V.~Thongchotchat, Y.~Kudo, Y.~Okada, and K.~Sato, ``Educational recommendation system utilizing learning styles: A systematic literature review,'' {\em IEEE Access}, vol.~11, pp.~8988--8999, 2023.

\bibitem{da2023systematic}
F.~L. da~Silva, B.~K. Slodkowski, K.~K.~A. da~Silva, and S.~C. Cazella, ``A systematic literature review on educational recommender systems for teaching and learning: research trends, limitations and opportunities,'' {\em Education and Information Technologies}, vol.~28, no.~3, pp.~3289--3328, 2023.

\bibitem{askarbekuly2021building}
N.~Askarbekuly, A.~Solovyov, E.~Lukyanchikova, D.~Pimenov, and M.~Mazzara, ``Building an educational product: constructive alignment and requirements engineering,'' in {\em Advances in Artificial Intelligence, Software and Systems Engineering: Proceedings of the AHFE 2021 Virtual Conferences on Human Factors in Software and Systems Engineering, Artificial Intelligence and Social Computing, and Energy, July 25-29, 2021, USA}, pp.~358--365, Springer, 2021.

\bibitem{kitchenham2004procedures}
B.~Kitchenham, ``Procedures for performing systematic reviews,'' {\em Keele, UK, Keele University}, vol.~33, no.~2004, pp.~1--26, 2004.

\bibitem{farina2022technologies}
M.~Farina, A.~Gorb, A.~Kruglov, and G.~Succi, ``Technologies for gqm-based metrics recommender systems: A systematic literature review,'' {\em IEEE Access}, 2022.

\bibitem{exercise_generation_ontology}
S.~Fischer, ``Course and exercise sequencing using metadata in adaptive hypermedia learning systems,'' vol.~1, p.~5–es, mar 2001.

\bibitem{pedagogical_framework}
S.~M. Parvez and G.~D. Blank, ``A pedagogical framework to integrate learning style into intelligent tutoring systems,'' {\em J. Comput. Sci. Coll.}, vol.~22, p.~183–189, jan 2007.

\bibitem{discussion_fora}
F.~Abel, I.~I. Bittencourt, E.~Costa, N.~Henze, D.~Krause, and J.~Vassileva, ``Recommendations in online discussion forums for e-learning systems,'' {\em IEEE Transactions on Learning Technologies}, vol.~3, no.~2, pp.~165--176, 2010.

\bibitem{blended7104183}
N.~Hoic-Bozic, M.~Holenko~Dlab, and V.~Mornar, ``Recommender system and web 2.0 tools to enhance a blended learning model,'' {\em IEEE Transactions on Education}, vol.~59, no.~1, pp.~39--44, 2016.

\bibitem{coding_literacy}
D.~S. Myers and N.~Chatlani, ``Implementing an adaptive tutorial system for coding literacy,'' {\em J. Comput. Sci. Coll.}, vol.~33, p.~260–267, dec 2017.

\bibitem{MLAcademia8693719}
H.~Samin and T.~Azim, ``Knowledge based recommender system for academia using machine learning: A case study on higher education landscape of pakistan,'' {\em IEEE Access}, vol.~7, pp.~67081--67093, 2019.

\bibitem{largescale}
K.~Dahdouh, A.~Dakkak, L.~Oughdir, and A.~Ibriz, ``Large-scale e-learning recommender system based on spark and hadoop,'' {\em Journal of Big Data}, vol.~6, no.~1, p.~2, 2019.

\bibitem{PariserumPerumal2019}
S.~Pariserum~Perumal, G.~Sannasi, and K.~Arputharaj, ``An intelligent fuzzy rule-based e-learning recommendation system for dynamic user interests,'' {\em The Journal of Supercomputing}, vol.~75, no.~8, pp.~5145--5160, 2019.

\bibitem{scratch8651403}
J.~Cárdenas-Cobo, A.~Puris, P.~Novoa-Hernández, J.~A. Galindo, and D.~Benavides, ``Recommender systems and scratch: An integrated approach for enhancing computer programming learning,'' {\em IEEE Transactions on Learning Technologies}, vol.~13, no.~2, pp.~387--403, 2020.

\bibitem{RecommendActions9249379}
N.~Yanes, A.~M. Mostafa, M.~Ezz, and S.~N. Almuayqil, ``A machine learning-based recommender system for improving students learning experiences,'' {\em IEEE Access}, vol.~8, pp.~201218--201235, 2020.

\bibitem{SubjectEnrollment9226409}
A.~J. Fernández-García, R.~Rodríguez-Echeverría, J.~C. Preciado, J.~M.~C. Manzano, and F.~Sánchez-Figueroa, ``Creating a recommender system to support higher education students in the subject enrollment decision,'' {\em IEEE Access}, vol.~8, pp.~189069--189088, 2020.

\bibitem{AR_DL_computational_thinking}
P.-H. Lin and S.-Y. Chen, ``Design and evaluation of a deep learning recommendation based augmented reality system for teaching programming and computational thinking,'' {\em IEEE Access}, vol.~8, pp.~45689--45699, 2020.

\bibitem{userPreferencesUsingHybridOptimization}
N.~Vedavathi and K.~M. Anil~Kumar, ``An efficient e-learning recommendation system for user preferences using hybrid optimization algorithm,'' {\em Soft Computing}, vol.~25, no.~14, pp.~9377--9388, 2021.

\bibitem{ersdoDynamicOntology}
M.~Amane, K.~Aissaoui, and M.~Berrada, ``Ersdo: E-learning recommender system based on dynamic ontology,'' vol.~27, p.~7549–7561, jul 2022.

\bibitem{theatre_context}
M.~Belver, A.~Manjarrés, A.~Barbarelli, and S.~Pickin, ``Requirements elicitation based on psycho-pedagogical theatre for context-sensitive affective educational recommender systems,'' {\em IEEE Access}, vol.~11, pp.~76284--76299, 2023.

\bibitem{recommend_review_materials}
F.~Okubo, T.~Shiino, T.~Minematsu, Y.~Taniguchi, and A.~Shimada, ``Adaptive learning support system based on automatic recommendation of personalized review materials,'' {\em IEEE Transactions on Learning Technologies}, vol.~16, no.~1, pp.~92--105, 2023.

\bibitem{game_vs_its}
Y.~Long and V.~Aleven, ``Educational game and intelligent tutoring system: A classroom study and comparative design analysis,'' {\em ACM Trans. Comput.-Hum. Interact.}, vol.~24, apr 2017.

\bibitem{usercentric7994718}
S.~Fazeli, H.~Drachsler, M.~Bitter-Rijpkema, F.~Brouns, W.~v.~d. Vegt, and P.~B. Sloep, ``User-centric evaluation of recommender systems in social learning platforms: Accuracy is just the tip of the iceberg,'' {\em IEEE Transactions on Learning Technologies}, vol.~11, no.~3, pp.~294--306, 2018.

\bibitem{children9956839}
M.~Moradi, K.~R. Fard, and M.~Y. Akhlaqi, ``A recommender system method for children’s education using mobile technology,'' {\em IEEE Access}, vol.~10, pp.~123679--123696, 2022.

\bibitem{usageContextBased6980102}
K.~Niemann and M.~Wolpers, ``Creating usage context-based object similarities to boost recommender systems in technology enhanced learning,'' {\em IEEE Transactions on Learning Technologies}, vol.~8, no.~3, pp.~274--285, 2015.

\bibitem{emotion9434237}
M.~Bustos~López, G.~Alor-Hernández, J.~L. Sánchez-Cervantes, M.~A. Paredes-Valverde, M.~d.~P. Salas-Zárate, and T.~Bickmore, ``Edurecomsys: An educational resource recommender system based on collaborative filtering and emotion detection,'' {\em Interacting with Computers}, vol.~32, no.~1, pp.~407--432, 2020.

\bibitem{ontology10}
J.~Joy, N.~S. Raj, and R.~V.~G., ``Ontology-based e-learning content recommender system for addressing the pure cold-start problem,'' vol.~13, no.~3, 2021.

\bibitem{SocialRetrieval2018}
C.~K. Pereira, F.~Campos, V.~Ströele, J.~M.~N. David, and R.~Braga, ``Broad-rsi – educational recommender system using social networks interactions and linked data,'' {\em Journal of Internet Services and Applications}, vol.~9, p.~7, 03 2018.

\bibitem{fuzzyTree7094243}
D.~Wu, J.~Lu, and G.~Zhang, ``A fuzzy tree matching-based personalized e-learning recommender system,'' {\em IEEE Transactions on Fuzzy Systems}, vol.~23, no.~6, pp.~2412--2426, 2015.

\bibitem{contexual_bandits}
W.~Intayoad, C.~Kamyod, and P.~Temdee, ``Reinforcement learning based on contextual bandits for personalized online learning recommendation systems,'' {\em Wireless Personal Communications}, vol.~115, no.~4, pp.~2917--2932, 2020.

\bibitem{heterogenous_evolution_network}
S.~Liu, S.~Liu, Z.~Yang, J.~Sun, X.~Shen, Q.~Li, R.~Zou, and S.~Du, ``Heterogeneous evolution network embedding with temporal extension for intelligent tutoring systems,'' {\em ACM Trans. Inf. Syst.}, vol.~42, nov 2023.

\bibitem{Semantic7272748}
P.~Montuschi, F.~Lamberti, V.~Gatteschi, and C.~Demartini, ``A semantic recommender system for adaptive learning,'' {\em IT Professional}, vol.~17, no.~5, pp.~50--58, 2015.

\bibitem{pomdp_representations}
J.~T. Folsom-Kovarik, G.~Sukthankar, and S.~Schatz, ``Tractable pomdp representations for intelligent tutoring systems,'' vol.~4, apr 2013.

\bibitem{lukyanchikova2023case}
E.~Lukyanchikova, N.~Askarbekuly, H.~Aslam, and M.~Mazzara, ``A case study on applications of the hook model in software products,'' {\em Software}, vol.~2, no.~2, pp.~292--309, 2023.

\bibitem{fabijan2017evolution}
A.~Fabijan, P.~Dmitriev, H.~H. Olsson, and J.~Bosch, ``The evolution of continuous experimentation in software product development: from data to a data-driven organization at scale,'' in {\em 2017 IEEE/ACM 39th International Conference on Software Engineering (ICSE)}, pp.~770--780, IEEE, 2017.

\bibitem{chi2014icap}
M.~T. Chi and R.~Wylie, ``The icap framework: Linking cognitive engagement to active learning outcomes,'' {\em Educational psychologist}, vol.~49, no.~4, pp.~219--243, 2014.

\bibitem{assessment_challenges}
R.~Ajjawi, J.~Tai, T.~L. Huu~Nghia, D.~Boud, L.~Johnson, and C.-J. Patrick, ``Aligning assessment with the needs of work-integrated learning: The challenges of authentic assessment in a complex context,'' {\em Assessment \& Evaluation in Higher Education}, vol.~45, no.~2, pp.~304--316, 2020.

\bibitem{zhao2019tackling}
Y.~Zhao, M.~Wehmeyer, J.~Basham, and D.~Hansen, ``Tackling the wicked problem of measuring what matters: Framing the questions,'' {\em ECNU Review of Education}, vol.~2, no.~3, pp.~262--278, 2019.

\bibitem{picard2004affective}
R.~W. Picard, S.~Papert, W.~Bender, B.~Blumberg, C.~Breazeal, D.~Cavallo, T.~Machover, M.~Resnick, D.~Roy, and C.~Strohecker, ``Affective learning—a manifesto,'' {\em BT technology journal}, vol.~22, no.~4, pp.~253--269, 2004.

\end{thebibliography}

\appendices

\section{Appendix A: Quality Assessment Score}
\label{appendix:quality_scores}
In addition to analysing the papers, we also attempted to assess the overall quality of papers in regards to clarity of motivation behind each paper, description of the proposed solution, and presence of evaluation section. We also gave each paper a subjective mark of how relevant it is to the purposes of our review.

We used the following questions to establish the quality score for each paper:

1) Was the motivation and intended use case for the development of the recommender system clearly specified?\\
-   1 point if clearly stated;\\
-   0.5 points if provided, but could be further elaborated;\\
-   0 points if hard to identify or not mentioned;\\

2) Are the user model, domain model, recommendation algorithm, and evaluation procedure clearly described and reproducible?\\
- 1 point if possible to replicate;\\
- 0.5 points if described , but difficult to replicate because of lack of details;\\
- 0 points if not described or not clear;\\

3) Was the proposed recommender system objectively evaluated?\\
-   1 point if conducted a fair and unbiased review of their algorithm or performed a critical analysis of its results;\\
-   0.5 points if performed an analysis, but partially biased or not clear or not critical enough;\\
-   0 points if did not conduct a fair and unbiased analysis or did not critically analyse the results.\\

4) Was the paper useful in regards to addressing the problem of assessing learning outcomes in recommender systems.\\
-   1 point if the paper directly addresses the assessment of learning outcomes in RS;\\
-   0.5 points if the paper is partially or indirectly addresses the assessment of learning outcomes in RS;\\
-   0 points if the paper is not relevant to the assessment of learning outcomes in RS.\\

The maximum possible score is 4 points. After calculating the score for each of the analyzed papers, we also calucated the average score for all papers and for each of the three databases (IEEE, ACM, and Springer).

\begin{figure}[h]
    \centering
    \includegraphics[width=0.45\textwidth]{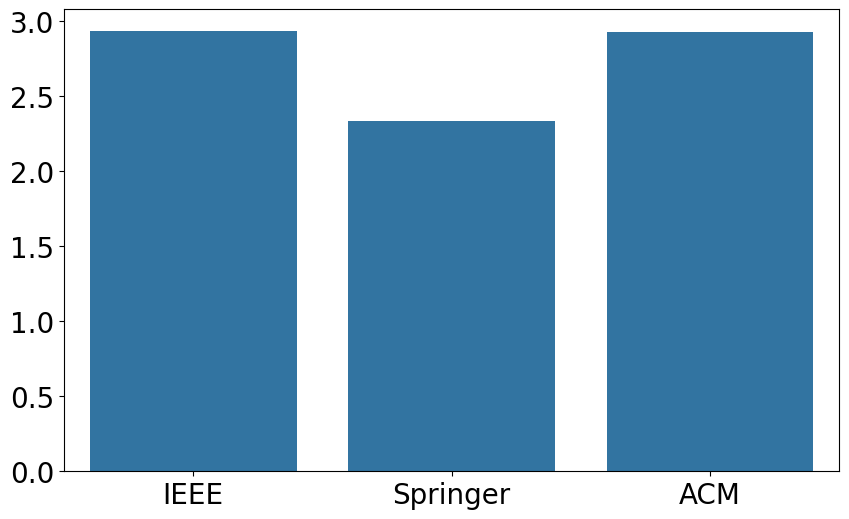}
    \caption{Average quality score by database}
    \label{fig:quality_by_source}
\end{figure}

The average quality score for papers is 2.804 out of 4.0. If we examine the average quality score by source (Fig. \ref{fig:quality_by_source}), we can see that IEEE and ACM Digital Library have a similar score, while Springer is slightly below.

Most importantly, this examination confirmed the overall high quality of the selection. For more details to see the quality score for each paper, please refer to our repository \footnote{\url{https://github.com/nurlingo/slr-ers/blob/main/data.csv}}.

\section{Descriptive summaries of the selected papers}
\label{appendix:summaries}
This appendix contains a short summary for each of the reviewed papers. For conveniece, we present the summaries in the same order as the works are referenced in the SLR.

% na43. Course and Exercise Sequencing Using Metadata in Adaptive Hypermedia Learning Systems.pdf

Fischer \cite{exercise_generation_ontology} used a LOM ontology to automatically generate assessment exercises based on the relations between concepts. The evaluation was a questionnaire to students, however not much detail is provided.

%na44. A pedagogical framework to integrate learning style into intelligent tutoring systems.pdf
Parvez and Blank \cite{pedagogical_framework} proposed a pedagogical framework to incorporate learning style when personalizing feedback for programming tasks. It is one of the early works advocating the importance of learning style.

% NI 3. Recommendations in Online Discussion Forums for E-Learning Systems.pdf

Abel et al \cite{discussion_fora} recommend threads in a discussion forum. They developed a loosely coupled recommendation module that can be integrated into a discussion forum of an e-learning system. They proceeded to ask how much data is required for collaborative filtering, what type of user feedback to consider, and whether user preferences change over time. At the end, the authors propose an interesting algorithm, where a different recommendation strategy is used based on how much data is available for the learner. They evaluate their system using standard precision recall metrics and an undisclosed internal dataset.

% 443. Recommender System and Web 2.0 Tools to Enhance a Blended Learning Model

Hoic-Bozic et al \cite{blended7104183} use a RS in the context of a university course to recommend optional collaborative activities and tools. They evaluate the system by comparing the formal assessment results between the years.

% na46. Implementing an Adaptive Tutorial System for Coding Literacy
Myers and Chatlani \cite{coding_literacy} uses Category-based Intrinsic Motivation algorithm to generate personalized learning content for coding literacy tasks. The learner model keeps track of user's correct and incorrect answers, and then uses a logistic function to calculate a learning score for each skill, reflecting the user's current understanding level.

% 446. Knowledge Based Recommender System for Academia Using Machine Learning: A Case Study on Higher Education Landscape of Pakistan

Samin and Azim \cite{MLAcademia8693719} use generative machine learning to represent user and domain through a probabilistic topic model, and then calculate the distance between topic models to make recommendations. The advantage of such approach is that the recommendations can be made on the basis of a querry in natural language.

% as12. Large‑scale e‑learning recommender system based on Spark and Hadoop

Dahdouh et al \cite{largescale} use association rules mining to recommend courses. The main idea behind the method is to find frequent combinations of courses that go together. The method is quite straightforward and only requires course enrollments history. The authors concentrate on the architecture and system implementation detail, however there is practically no evaluation for the method itself as they only compare the time performance of various algorithms that do the mining.

% as11. An intelligent fuzzy rule‑based e‑learning recommendation system for dynamic user interests

Pariserum Perumal et al \cite{PariserumPerumal2019} recommends an e-learning content by finding patterns in learner behavior. In particular, they review the history of visited items by topics, and then conclude the learner's interests based on that. The paper attempts to take into account the "dynamic user interests" by using fuzzy logic and splitting the history into windows, instead of treating as one whole. This can be useful in an e-learning context, where a user's interests might evolve as they learn new things. For evaluation, they measure the relevance through the Harmonic mean. However, the learning outcomes and their achievement is not touched in the paper at all.

% 449. Recommender Systems and Scratch - An Integrated Approach for Enhancing Computer Programming Learning.pdf

Cardenas-Cobo et al \cite{scratch8651403} recommend Scratch programming tasks to university students that are appropriate to their level.    They use traditional CF approach, where each student has the set of solved exercises with the corresponding ratings tuple (taste, complexity).    Based on the ratings three types of similarity between users are computed, and then a set of similar users is produced. The study implements a thorough evaluation methodology, where both a qualitative questionnaire (including RS performance, user-centric effects, effects of learning)  and a quantitative academic performance data are used.

% 447. Recommender Systems and Scratch: An Integrated Approach for Enhancing Computer Programming Learning

Yanes et al \cite{RecommendActions9249379} recommend course improvement actions to instructors based on the course description, learning outcomes and assessment results. They frame it as a classification problem, and compare several ML algorithms to find the significant features and best performing method. Despite commenting on the importance of the outcome-based education, they do not measure how their recommendations affects the achievement of COs.

% 448. Creating a Recommender System to Support Higher Education Students in the Subject Enrollment Decision.pdf

Fernández-García et al \cite{SubjectEnrollment9226409} describe the full procedure of creating an ML-based RS to help students choose academic subjects. The work also serves as a great tutorial to pre-processing educational data, and describes all taken steps in a clear reproducible way. Importantly, it demonstrates how feature engineering and resampling can partially compensate the imbalance and scarcity in data. However, the recommender system aspect is not significantly highlighted, and the paper is more of a study on ML classification and data pre-processing.

% NI 28. Design and Evaluation of a Deep Learning Recommendation Based Augmented Reality System for Teaching Programming and Computational Thinking
Lin and Chen \cite{AR_DL_computational_thinking} integrated a deep learning-based recommender into a programming course to recommend course activities. They split student into a control and experiment groups and conducted a pre- and post-tests to measure the effect of using the recommender on the learning outcomes of the experiment group.

% as17. An efficient e-learning recommendation system for user preferences using hybrid optimization algorithm

Vedavathi and Anil Kumar \cite{userPreferencesUsingHybridOptimization} recommend web resources to students in university. They used a Deep Recurrent Neural Network and an improved whale optimization algorithm to classify the e-learners and order their ratings in sequence.

% as19. ERSDO: E-Learning Recommender System Based on Dynamic Ontology
Amane et al \cite{ersdoDynamicOntology} recommend courses to university students using dynamic ontologies to represent the learners, courses, and links between them. Among other properties, they take into account languages preferences and learning style. Their method is an interesting hybrid where knowledge-based approach is used to conduct clustering of similar learners. The dynamicity in their approach means that they update the ontologies on regular basis instead of calculating them once in the very beginning.

% NI 55. Requirements Elicitation Based on Psycho-Pedagogical Theatre for Context-Sensitive Affective Educational Recommender Systems.pdf

Belver et al \cite{theatre_context} propose a requirements elicitation technique in the form of theatre that can help to extract a broader learning context including affective dimensions. Then they design an ontology that encapsulates these learner and context properties including learning style, mood, and even weather. They make recommendations by finding the most similar learner context accross these many dimensions.

% NI 60 Adaptive Learning Support System Based on Automatic Recommendation of Personalized Review Materials

Okubo et al \cite{recommend_review_materials} recommend revision materials to university students. The recommendations are based on the quiz results, whereby the system suggests lecture slides that are semantically similar to the questions where the learner made mistakes. As their evaluation, they split students into control and experiment groups and used a pre and post-test to see which group performs better on average.

% na45.    Educational Game and Intelligent Tutoring System: A Classroom Study and Comparative Design Analysis
Long and Aleven \cite{game_vs_its} describe an ITS for learning algebraic equations, which uses Bayesian Knowledge Tracing (BKT) to model a student's mastery of skills over time and select appropriate exercises. They compare the ITS against a popular educational game from the same domain in a split study, and then evaluate the learning outcomes and learners' enjoyment for each system. Interestingly, their results indicate that the enjoyment does not necessarily correlate with the actual learning. However, it does lead to greater engagement.

% 444. User-centric Evaluation of Recommender Systems in Social Learning Platforms: Accuracy is Just the Tip of the Iceberg.pdf

Fazeli et al \cite{usercentric7994718} note that evaluating RS from the point of view of accuracy is not sufficient, and it should be supplemented with a qualitate set of user-centric metrics. They also run an experiment to demonstrate that graph-based CF can outperform user-based CF in regards to these qualitative metrics. The work is important, as it questions the validity of precision as the default metric to measure effectiveness of RS in the context of education.

% 451. A Recommender System Method for Children’s Education Using Mobile Technology

Moradi et al \cite{children9956839} recommend educational resources to children using mobile technology and by observing the group dynamics. They model interaction between children as a directed acyclic graph (DAG), and assign subject topics to each child. Then based on the interaction data, they suggest educational resource in the topics which were labelled as 'weak'. They evaluate the method using simulation to see how fast the correct information will disseminate in the DAG. Importantly, they recognize that children learn from each other, and by suggest that by influencing the most active nodes in the graph an educator can improve the learning for the whole group.

% 442. Creating Usage Context-based Object Similarities to Boost Recommender Systems in Technology Enhanced Learning.pdf

Niemann and Wolpers \cite{usageContextBased6980102} propose a way to calculate similarity between learning objects implicitly based on what items were used together in one session. They argue that this can be an alternative to the content-based approach when metadata on the learning objects is not available or difficult to gather. The main idea behind their approach is that "users' knowledge and context is inherent in their activities", therefore based on the co-usage of certain objects, one can derive the correlation between them.

% 450. EduRecomSys - An Educational Resource Recommender System Based on Collaborative Filtering and Emotion Detection.pdf

Bustos López et al \cite{emotion9434237} suggest a CF-based recommender that incorporates emotion detection results into the ratings, and compare it to a traditional CF algorithm. The evaluation includes user-centric questionnaire in addition to F1 metrics, which provides us with both a quantitative and qualitative perspectives. They also make a weighted matrix to compare the features with other similar systems. Though it is not clear why a system with most features should necessarily be superior, however, having the set of commonly-occurring features is of value in itself.

% aj9. Ontology-based E-learning Content Recommender System for Addressing the Pure Cold-start Problem.pdf

Joy et al \cite{ontology10} propose a generic ontology for e-learning that combines the learner, learning objects, and learner's path into a single model. The ontology allows to recommend resources even for new users as the article's main goal is to solve the pure cold start problem. The learner model includes personal information and learning styles, however there are no details on the privacy aspect of the user data. The evaluation is done by gathering real users' satisfaction with the recommendations on 1 to 5 scale.

% s9. BROAD-RSI – educational recommender system using social networks interactions and linked data.pdf

Pereira et al \cite{SocialRetrieval2018} propose to extract learner data from social network profiles. They argue that social networks are a much richer source of data, rather than traditional LMS. The authors use entities, concepts, and keywords detection and domain ontologies to extract user interests from Facebook posts and groups. They let users confirm the correctness of the extracted information. The resources are recommended from several sources including Open University learning objects, DBpedia and Youtube. To evaluate the recommendations they conducted a case study and used questionnaires to ask the participants about the relevance of extracted interest and recommended items.

% 440. A Fuzzy Tree Matching-based Personalized e- Learning Recommender System.pdf

Wu et al \cite{fuzzyTree7094243} propose an approach where both learning activities and learner profiles are represented as fuzzy trees. Various techniques are applied to calculate similarity be between learning activities based on categories assigned to them. The fuzzy tree combinations are used to derive the fuzzy tree of all categories learned by a learner. Moreover, the model facilitates for a manual specification of prerequisite relation between activities and also automatic generation of precedence relations from the sequences in learners' histories. The learner profile contains learner preference scores for available categories in the category tree, and the scores are used to calculate similarity between users and to match a learner with an activity. A hybrid recommendation method is used, including the knowledge of relations between the items and users' preferences and learning goals, and the use of collaborative technique that takes into account ratings by similar users.

% S35. Reinforcement Learning Based on Contextual Bandits for Personalized Online Learning Recommendation Systems.pdf
Intayod et al \cite{contexual_bandits} developed a reinforcement learning agent to recommend learning objects to students. As their algorithm, they suggested a modification of a multi-arm bandit. Their aim was to maximize clickthrough rate.

% na48. Heterogeneous Evolution Network Embedding with Temporal Extension for Intelligent Tutoring Systems
Liu et al \cite{heterogenous_evolution_network} proposed a method to model the learners, resources, exercises, and knowledge concepts within a single heterogeneous graph. The graph includes the learner interactions with resources and exercises, and maps the exercises to knolwedge concepts. After a few transformations it can serve as the training data for deep learning methods.

% 453. A Semantic Recommender System for Adaptive Learning.pdf

Montuschi et al \cite{Semantic7272748} help job seekers to acquire missing skills for a particular vacancy. The identify a gap between the user and his desired job, and then formulate it as desired learning outcomes (LOs). Then they use a semantic taxonomy to map the LOs to the educational resources. The system is a classic knowledge-based recommender system, however authors themselves do not refer to it as such. Interestingly, they allow users to participate in mapping resources to LOs, thus using a form of crowdsourcing to augment the system's knowledgebase.

% na39. Tractable POMDP Representations for Intelligent Tutoring Systems.pdf
% ITS

Folsom-Kovarik et al \cite{pomdp_representations} proposed an intelligent tutoring system based on Partially Observable Markov Decision Process. The system models learner as a set of knowledge states each representing a desired learning outcome that learner needs to cover. Then the system suggests corrective actions to increase the probability of each state being covered and not leaving gaps.

\begin{IEEEbiographynophoto}{Nursultan Askarbekuly}
Nursultan Askarbekuly is a PhD student at the Institute of Software Engineering, Innopolis University and at the Faculty of Organizational Sciences, University of Belgrade. His main research interest is data-driven educational technology, empirical software engineering, and product development.
\end{IEEEbiographynophoto}

\begin{IEEEbiographynophoto}{Ivan Luković}
Ivan Luković is a Full Professor at University of Belgrade, Faculty of Organizational Sciences. His main research interests are focused on Theory of data models, System design, particularly logical and physical database design, and development and usage of MDSD / CASE tools in Software Engineering and System Design.
\end{IEEEbiographynophoto}

\end{document}